\documentclass[journal]{IEEEtran}
\ifCLASSINFOpdf
\else
\fi

\usepackage{float}

\usepackage{cite}
\ifCLASSINFOpdf
\else

\usepackage[dvips]{graphicx}

\DeclareGraphicsExtensions{.eps}
\fi
\usepackage[usenames]{color} 
\usepackage{soul}
\usepackage{times}
\usepackage{amsmath}
\usepackage{epsfig}
\usepackage{colortbl}
\definecolor{light-gray}{gray}{0.95}
\definecolor{gray}{gray}{0.6}
\definecolor{dark-gray}{gray}{0.35}
\usepackage{multirow}
\usepackage[table,dvipsnames]{xcolor}
\usepackage{algorithm,algpseudocode,float}
\usepackage{amssymb}
\usepackage{stfloats}
\usepackage{multirow}
\emergencystretch=\maxdimen
\begin{document}
%
\title{Experimental Analysis on Variations and Accuracy of Crosstalk in Trench-Assisted Multi-core Fibers}
%
%
%

\author{Hui Yuan,~\IEEEmembership{} Alessandro Ottino, Yunnuo Xu, Arsalan Saljoghei, Tetsuya Hayashi, Tetsuya Nakanishi, Eric Sillekens, Lidia Galdino,  Polina Bayvel, Zhixin Liu, and Georgios Zervas
\thanks{H. Yuan, A. Ottino, Y. Xu, A. Saljoghei, E. Sillekens, L. Galdino, P. Bayvel, Z. Liu, and G. Zervas are with the Optical Networks Group, University College London, London
WC1E 7JE, UK (e-mail: h.yuan@ucl.ac.uk; alessandro.ottino.16@ucl.ac.uk;  yunnuo.xu.18@ucl.ac.uk; a.saljoghei@ucl.ac.uk; e.sillekens@ucl.ac.uk; l.galdino@ucl.ac.uk; p.bayvel@ucl.ac.uk; zhixin.liu@ucl.ac.uk; g.zervas@ucl.ac.uk).}
\thanks{T. Hayashi and T. Nakanishi are with Optical Communications Laboratory, Sumitomo Electric Industries, Ltd., 1, Taya-cho, Sakae-ku, Yokohama 244-8588, Japan (e-mail: t-hayashi@sei.co.jp; nakanishi-tetsuya@sei.co.jp).}
\thanks{Manuscript received July XX, 2018; }}

\maketitle

\begin{abstract}

Space division multiplexing using multi-core fiber (MCF) is a promising solution to cope with the capacity crunch in standard single-mode fiber based optical communication systems. Nevertheless, the achievable capacity of MCF is limited by inter-core crosstalk (IC-XT). Many existing researches treat IC-XT as a static interference, however, recent research shows that IC-XT varies with time, wavelength and baud rate. This inherent stochastic feature requires a comprehensive characterization of the behaviour of MCF to its application in practical transmission systems and the theoretical understanding of IC-XT phenomenon. In this paper, we experimentally investigate the IC-XT behaviour of an 8-core trench-assisted MCF in a temperature-controlled environment, using popular modulation formats. We compare the measured results with the theoretical prediction to validate the analytical IC-XT models previously developed. Moreover, we explore the effects of the measurement configurations on the IC-XT accuracy and present an analysis on the IC-XT step distribution. Our results indicate that a number of transmission parameters have significant influence on the strength and volatility of IC-XT. Moreover, the averaging time of the power meter and the observation time window can affect the value of the observed IC-XT, the degrees of the effects vary with the type of the source signals. 

\end{abstract}

\begin{IEEEkeywords}
Multi-core fiber, crosstalk, transmission parameters, accuracy, IC-XT step distribution.
\end{IEEEkeywords}

%
\IEEEpeerreviewmaketitle

\section{Introduction}
\label{sec:intro}

\IEEEPARstart{I}{n} the past decades, optical transmission systems have been developing rapidly \cite{Kachris_2012}. The capacity crunch in standard single-mode fiber (SMF) imposes a limit in further scaling the capacity of optical communication systems \cite{Richardson_2013}. To meet the capacity demands in the next generation optical communication systems, space division multiplexing (SDM) technologies employing multi-core fiber (MCF) were proposed, which show advantages on SMF in terms of capacity, wiring complexity and front panel density, with a simultaneous reduction in power consumption and cost. This makes it a popular solution for different optical communication systems including long-haul communication \cite{Puttnam_2016} and short-reach data center networks (DCNs) \cite{Liu_2016},\cite{Yuan_2018}. Recent research showed that a MCF-based network can save hardware by sharing transceiver digital signal processing (DSP), compared to the SMF-based network  \cite{Klaus_2017}. In another research, silicon photonic (SiP) on-board transceivers coupled with MCFs can be used to increase the front panel density of the transmission system while offering better space management \cite{Hayashi_2017}. Moreover, optical switches that employ MCFs can route signals from multiple cores together to support purely MCF-based DCN links with SDM switching \cite{Mulvad_2017}.

Promising as it is, the application of the MCFs in optical networks may be limited by the inter-core crosstalk (IC-XT) between the adjacent cores, which can reduce the optical signal-to-noise ratio (OSNR), and subsequently, the system performance and power budget \cite{Puttnam_2016}. To date, many investigations on MCF-based networks consider IC-XT as a static value that can be calculated from the existing analytical models~\cite{Muhammad_2014,Yao_2017,Yuan_2018a,Yang_2019,Tode_2017,Yao_2018}. Nevertheless, it was shown in \cite{Rademacher_2017, Hayashi_2014} that IC-XT can vary by up to 15~dB over 60 minutes observation period. Consequently, the IC-XT fluctuations may enforce the system to set a higher OSNR margin to ensure sufficient system performance. The IC-XT variation can be attributed to the longitudinally varying perturbation along the MCF, such as macro bending, twists and structural fluctuations \cite{Hayashi_2011a}. Several mathematical models have been developed to analyze the IC-XT behavior, including the fluctuations in homogeneous MCFs \cite{Hayashi_2011a, Luís_2016, Cartaxo_2016, Alves_2016}. Among which, the model based on a Brownian motion for time dependent IC-XT greatly fits the experimental results \cite{Alves_2016}. Moreover, it has been theoretically predicted that IC-XT and its fluctuations are associated with the operating wavelength \cite{Ye_2016} and skew between the adjacent cores \cite{Rademacher_2017}.

The symbol rate and modulation format of the stimulating optical signals can also influence the IC-XT fluctuations. Experimental results from \cite{Rademacher_2017} show that the fluctuations can be decreased by either increasing the signalling bandwidth of intensity modulated signals or by adopting the phase modulation methods \cite{Rademacher_2017}. This improvement was obtained by the reduction or elimination of the power of the residual optical carrier \cite{Hayashi_2018}. However, comparisons between different intensity modulated formats have not been studied. In addition, for phase modulated formats, the effects from quadrature phase shift keying (QPSK) and binary phase shift keying (BPSK) with various symbol rates on IC-XT fluctuations have been theoretically and experimentally investigated in \cite{Rademacher_2017}, respectively. However, for higher order phase modulated format, only dual-polarization 16-quadrature amplitude modulation (DP-16QAM) in a single wavelength, at the signal rate of 24.5 GBaud, has been experimentally demonstrated.

Additionally, temperature changes in DCNs \cite{Clark_2018} and the temperature fluctuations over silica fibers can cause length change by $4.1\times10^{-7}$ mK$^{-1}$ and refractive index variation by $1.1\times10^{-5}$ K$^{-1}$ \cite{Slavík_2016}. These changes can alter the latency/skew as well as dispersion in the fiber. Given the characteristics of MCFs, it is clear that temperature fluctuations can have a similar impact on these fibers, e.g. higher temperature leads to longer skew\cite{Puttnam_2018}, provided that evolution of IC-XT across the MCF is dependent on skew and structural fluctuations along the fiber. Therefore, it is believed that temperature variations across MCF can impact IC-XT characteristics. However, to the best of the authors' knowledge, there have been limited studies on the impact of temperature variation on IC-XT \cite{Hayashi_2014}. Moreover, most of the studies which shed some light on characteristic of IC-XT fluctuations were limited in determining IC-XT fluctuations at sub-Hz levels~\cite{Rademacher_2017, Luís_2015, Alves_2018, Alves_2019, Luís_2016}, except the work presented in \cite{Rademacher_2017a}, where a 5 kS/s oscilloscope was used. 

The aforementioned transmission parameters can internally affect the IC-XT behavior inside the MCF. However, in the measurement process in the lab, the value of the observed IC-XT can be externally affected by the configuration of the devices, e.g. averaging time of the power meter, and the observation time~\cite{Alves_2019}. In the existing studies, different averaging times have been used~\cite{Luís_2016,Luís_2015,Alves_2018,Alves_2019}, however, the effect of the averaging time on the observed IC-XT has not been investigated. Although, the time window effect has been analyzed in \cite{Alves_2019}, only the narrow band continuous-wave (CW) source light was experimentally demonstrated.
 
This paper extends our OFC publication \cite{Yuan_2019} by presenting a more comprehensive study of the behaviour of IC-XT in MCFs. The novel contributions from the work presented in this paper are listed below:
\begin{itemize}
\item The IC-XT variations using different advanced modulation formats, including QAM with various cardinalities and pulse amplitude modulation (PAM), and baud rates are studied.
\item The IC-XT was measured over periods up to 12 hours, for an ultra-wide spectral window spanning 250 nm (O-S-C-L bands), to attain a full picture of it.
\item For the first time, the impact of temperature and length of pseudo random binary sequence (PRBS) on IC-XT behaviour, is evaluated. 
\item A novel study of the impact of observation time and averaging time on the observed IC-XT, which is meaningful for IC-XT investigation in the lab and real world, is presented. 
\item A new model for IC-XT step distribution, which is effective for IC-XT induced by all the investigated sources, is proposed.
\end{itemize}

In addition, compared to the MCFs utilized in the existing research~\cite{Rademacher_2017, Alves_2019}, the fiber employed in this work is much shorter as it is specifically fabricated for short reach ($<$ 1~km) data center transmission. Also, the fiber type and core pitches are different, which may lead to different IC-XT behavior. Therefore, we have also done some similar analysis that have been done in~\cite{Rademacher_2017} with this new fiber. Furthermore, the results are compared to the theoretical estimations to verify the analytical fiber models.

The rest of the paper is structured as follows. A brief overview of static and dynamic IC-XT in MCF along with the related mathematical models are presented in Section \ref{sec:ICXT}. Section \ref{sec:set} illustrates three experimental scenarios and the corresponding setups. Section~\ref{sec:res} analyzes the numerical results in various aspects. Section~\ref{accu} explores the experimental setup effect on the accuracy of the observed IC-XT and analyzes the IC-XT step distribution. Section~\ref{sec:con} provides the concluding remarks.

\section{Static and Dynamic Inter-core XT in MCF}

\begin{figure*}[t]
	\centering
	\includegraphics[width = 1.95\columnwidth]{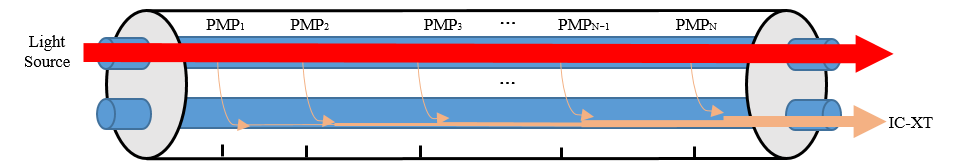}
	\caption{IC-XT generation process in MCF.}
	\label{fig:PMPs}
\end{figure*}

\label{sec:ICXT}
IC-XT is an unwanted interference inherent to MCFs, which can be considered as the power leakage from one core to the target core. It occurs at stochastically distributed discrete points along the fiber where the principal and IC-XT signals match in phase \cite{Hayashi_2011a}. The discrete points are called phase-matching points (PMPs), shown in Fig.~\ref{fig:PMPs}, where the total IC-XT can be approximated as the sum of the crosstalk contributions overall. According to \cite{Luís_2016}, the number and/or distribution of the PMPs can vary randomly with each core's characteristics, fiber structure and the conditions under which the fiber is placed. Especially, the bending radius, propagation constants and twist rate fluctuations that lead to the phase offset variance, will make IC-XT dynamic \cite{Hayashi_2010}.

\subsection{Static IC-XT}
\label{subsec:Sta}
In this paper, the term ``static IC-XT" will be used to represent the average IC-XT for a given time ($\ge$ 1 hour), which is approximately equal to the statistical mean IC-XT. In other words, static IC-XT can also be seen as the statistical mean IC-XT, which can be calculated by Eq.~(\ref{eq:meanXT}) \cite{Koshiba_2012}, \cite{Ye_2014} in a homogeneous MCF.

\begin{equation}
\label{eq:meanXT}
XT(\kappa ,{C_P}) = 2\frac{{{\kappa ^2}R}}{{\beta {C_P}}}L 
\end{equation}

In this equation, $R$, $\beta$, $C_p$ and $L$ represent the bending radius, propagation constant, core pitch between two adjacent cores and fiber length, respectively. $\kappa$ is the mode coupling coefficient and for TA-MCF it can be expressed as \cite{Ye_2016}:

\begin{equation}
\label{eq:ka}
\kappa  = \frac{\sqrt{\Delta _1}}{a} 
\frac{U_1^2}{V_1^3 K_1^2 \left( W_1 \right) } 
\sqrt{ \frac{\pi a \Gamma}{W_1 C_p} } 
e^{\left[ - \frac{W_1 C_p}{a} - 2\left( W_2 - W_1\right) \frac{w_t}{a} \right] }
\end{equation}

\noindent Where $a$ and $w_t$ stand for the core radius and trench width, respectively. ${U_1} = {a}{({k^2}n_1^2 - {\beta ^2})^{1/2}}$ and ${V_1} = k{a}{n_1}{(2{\Delta _1})^{1/2}}$. $K_1(W_1)$ is the modified Bessel function of the second kind with first order and ${W_1} = {a}{({\beta ^2} - {k^2}n_0^2)^{1/2}}$. $k = 2\pi/\lambda$ is the wave number, and $\lambda$ stands for the wavelength of the light. $\Gamma  = {W_1}/[{W_1} + ({W_2} - {W_1}){w_t}/{C_p}]$. ${W_2} = {(V_2^2 + W_1^2)^{1/2}}$, in which ${V_2} = k{a}{n_0}{(2|{\Delta _2}|)^{1/2}}$. $n_1$ and $n_0$ stand for the refractive index of core and cladding, respectively, while $\Delta _1$ is the difference between them. $\Delta _2$ denotes the refractive index difference between trench and cladding.

Mathematically, Eq.~(\ref{eq:meanXT}) indicates that if the core pitch, bending radius and fiber length are fixed, the static IC-XT is proportional to $\kappa^2$, where $\kappa$ is related to wavelength ($\lambda$). Therefore, IC-XT is wavelength dependent, i.e. the longer the wavelength, the higher the static IC-XT. Moreover, changes on the refractive index difference between the core and cladding or cladding and trench, e.g. due to temperature changes, can considerably alter the static IC-XT.

\subsection{Dynamic IC-XT}
\label{Dynamic}
For distinction, ``dynamic IC-XT" is used to define the IC-XT variation or fluctuation range (i.e. difference between the maximal and minimal IC-XT) for a long term ($\ge$ 1 hour). According to references~\cite{Luís_2016}, \cite{Alves_2016} and \cite{Rademacher_2017a}, in the frequency domain, time-dependent IC-XT between the excited core (m) and the target core (n) can be modeled as:
 
\begin{equation}
\label{eq:Amp}
A_{n}({z_l},\omega) = A_{n}(0,\omega) - jK\sum_{l=1}^Ne^{-j\left[\Phi_l(t)+{sz_l}\omega\right]}{A_{m}(z_{l-1},\omega)}
\end{equation}

\noindent Where $A_{n}(z_l,\omega)$ stands for the complex amplitude of the IC-XT signal in the target core (n) at the $l$-th PMP. $\omega$ and $t$, denote the angular frequency and time, respectively. $s$ is walk-off, which represents a delay for unit of length. $N\approx L{\gamma}/\pi$, is the number of PMPs, in which $\gamma$ denotes the twist rate. $\Phi(t)$ represents the time-dependent phase offset between every two adjacent PMPs, which randomly varies between 0 to 2$\pi$. $z_l$ stands for the distance of the $l$-th PMP along the fiber from the start of the fiber and $K$ is the discrete coupling coefficient between the two cores can be calculated by:

\begin{equation}
|K|  = \sqrt{\frac{{{2\pi\kappa^2 R}}}{{\beta{C_p}\gamma}}}
\end{equation}

\noindent When the IC-XT is adequately low ($A_{n}({z_l},\omega)\ll1$), $A_{n}(0,\omega) = 0$ and ${A_{m}(z_{l-1},\omega)}=1$, Eq.~(\ref{eq:Amp}) can be simplified as:
 
\begin{equation}
A_{n}(L,\omega) \approx - jK\sum_{l=1}^Ne^{-j\left[\Phi_l(t)+{sz_l}\omega\right]}
\end{equation}

According to the analysis in~\cite{Rademacher_2017}, if signals are modulated with low symbol rates or the skew between the two cores ($\approx sL$) is small, $\Phi$ will be much bigger than $sz_l\omega$ and thus, the IC-XT will have higher dependence on time (high variance). On the contrary, when signals are modulated with high symbol rates or the skew between the two cores is big, $sz_l\omega$ will dominantly affect the IC-XT variance and thus, the IC-XT will be more stable.

In addition, in terms of modulation format, if the power of the residual optical carrier is either reduced or eliminated, the IC-XT variance will decrease accordingly as the IC-XT of the signal light will be better averaged over the signal band \cite{Hayashi_2018}. Furthermore, PRBS pattern is also a parameter that should be considered. The reasons are a) PRBS pattern is an important parameter for bit error rate evaluation in transmission testbeds and research groups can easily use different PRBS patterns; b) PRBS pattern can change the properties of the power spectral density (PSD) of the modulated signal \cite{Rice_2016, Wiley}, which may potentially change the IC-XT. However, to the best of the authors' knowledge, there have been no study on the effect of PRBS pattern on IC-XT.

\section{Experimental Setup}
\label{sec:set}
\begin{figure*}[t]
	\centering
	\includegraphics[width=2\columnwidth]{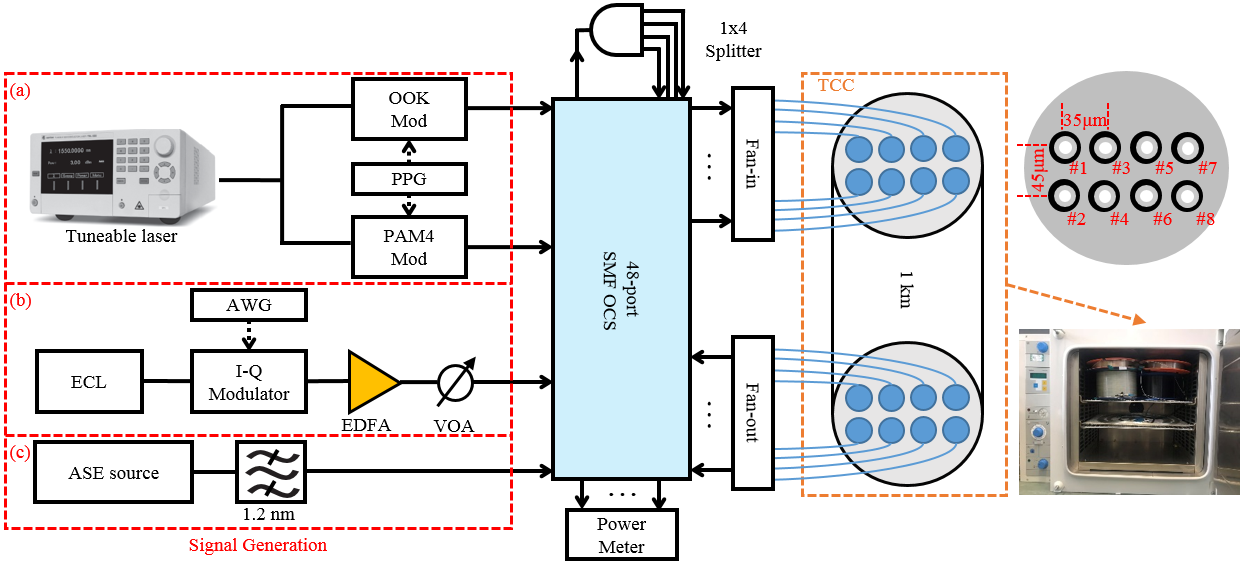}
	\caption{Experimental setup for measuring the time-dependent IC-XT and profile of the fabricated 8-core TA-MCF.}
	\label{fig:setup}
\end{figure*}

The cross sectional area of the 8-core TA-MCF employed in our experiment is shown on the right of Fig.~\ref{fig:setup}. This fiber had been previously showcased in  \cite{Hayashi_2017}, which has an 180 $\mu$m cladding diameter and a 0.17 m bending radius. The core pitch between two horizontal neighbouring cores is 35 $\mu$m and 45~$\mu$m between neighbours existing vertically.

The experimental setup used to measure the static and dynamic IC-XT characteristics of the TA-MCF is illustrated in Fig.~\ref{fig:setup}. As it can be seen, to fully study the IC-XT behaviour, various types of input stimulus were employed. In the first subsystem (a), in order to investigate the IC-XT characteristics of the MCF over an ultra-wide bandwidth, two tunable lasers, one for O-band (1260-1360 nm) with 500~kHz linewidth and one for S-, L- and C-band (1480-1630 nm) with 200~kHz linewidth, were used to generate a CW laser light, respectively. A Mach-Zehnder modulator (MZM) driven by an electrical modulation signal containing either on-off keying (OOK) or PAM-4 signals modulated the CW seed light. Both the OOK and PAM-4 signals were generated via a pulse pattern generator (PPG), which could operate with 10 or 25~GBaud signalling rates. Moreover, this PPG had a reconfigurable PRBS patterns with lengths of 2$^{i}$-1 (\emph{i} = 7, 9, 10, 11, 15, 20, 23 and 31). In the second subsystem (b), to evaluate the characteristics of the IC-XT induced by various phase modulated signals, a 100 kHz linewidth external cavity laser (ECL) operating at 1550 nm was modulated by a dual-polarization I-Q modulator driven by a 92 GS/s arbitrary waveform generator (AWG). The AWG was utilized to generate various \emph{m}-ary (\emph{m} = 4, 16, 64 and 256) dual-polarization QAM formats. Each of the QAM format could operate at various signalling rates ranging from 15-80~GBaud. The modulated optical signal after the I-Q modulator was then amplified by an erbium-doped fiber amplifier (EDFA), where its launch power was controlled by a variable optical attenuator (VOA). In the third subsystem (c), a C-band ASE source was generated and its output was passed through a band-pass filter to generate a broadband signal with a 1.2 nm bandwidth prior to launch into the MCF. 

In order to automate the IC-XT measurement routine, a 48-port optical circuit switch (OCS) was adopted to interconnect all the input ports of the MCF link via the fan-in device to the various signal sources over the SMF links. The switch enabled the generated signals to be sent into any group or individual core of the MCF via the fan-in device with/without passing through a 1x4 splitter, which enabled the replication of the same signal source. The 8-core TA-MCF used in this work had a total length of 1 km, and in order to control its temperature during the experimental procedure, this fiber was resided within a temperature controlled chamber (TCC), which could shift the temperature from 20 to 80 $^\circ$C. To measure the IC-XT level, the output of the MCF was fanned-out into eight SMF links, each connected to the input of an 8-port high speed optical power meter operating at 40 Hz, capable of detecting power levels from -80 to +10~dBm~\cite{Santec}. The observed IC-XT is the ratio of output power of the crosstalk core to output power of the excited core. In this setup, the switch, splitter and fan-in/out device approximately had 1 dB, 6 dB and 3~dB of insertion loss, respectively. The crosstalk of the switch ports used in the experiment was less than -90 dB as the sensitivity of power meter used was -90~dBm and there was no light detected. It should also be mentioned that the launch power for all various scenarios was set to 4~dBm at the input port of the optical switch, and thus, launch powers through the MCF were not sufficient to cause any nonlinearities. 

\section{Results and Discussion}
\label{sec:res}
With the setup illustrated in Section \ref{sec:set}, we have executed numerous of experiments in various environments to explore the behaviour of the static and dynamic IC-XT. The variables include temperature, wavelength, PRBS length, light source, modulation format, baud rate and number of excited cores. The obtained results are shown and analyzed in the following subsections.

\begin{figure*}[t]
\begin{minipage}[t]{\columnwidth}
\centering
\includegraphics[width=0.99\columnwidth]{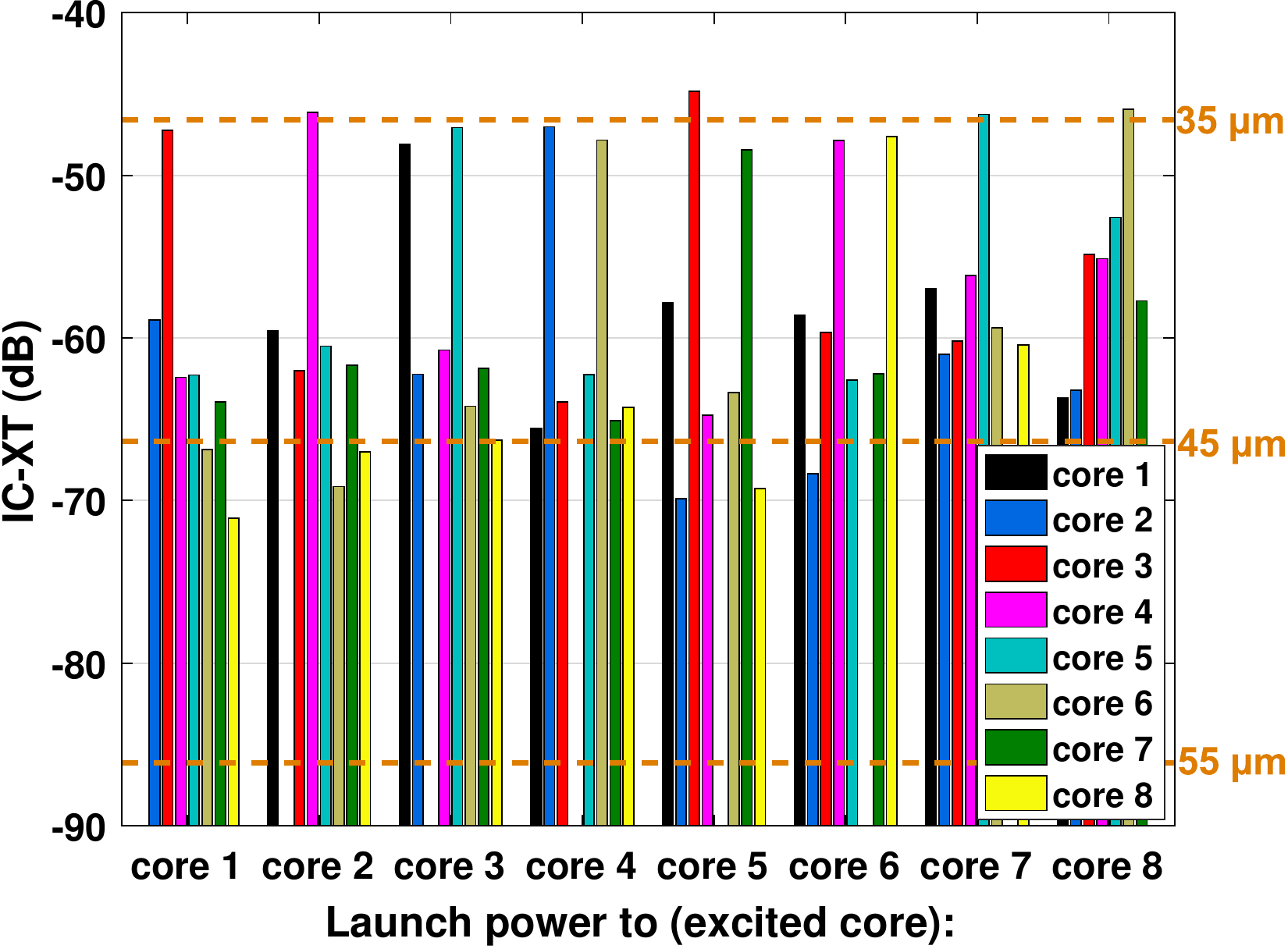}
\caption{ Measured and analytical pairwise IC-XT between cores}
\label{fig:pairwise}
\end{minipage}%
\hfill 
\begin{minipage}[t]{\columnwidth}
\centering
\includegraphics[width=0.99\columnwidth]{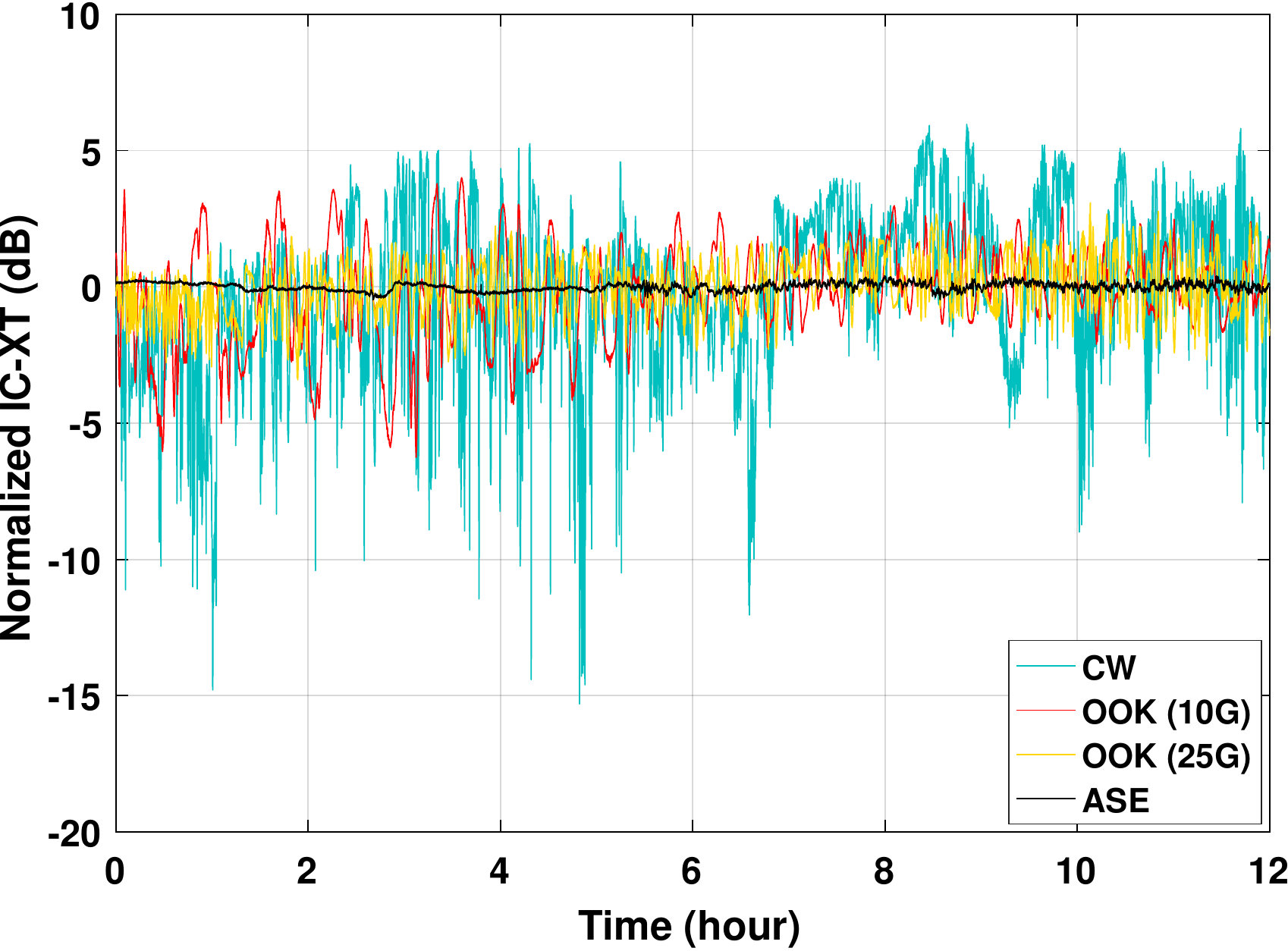}
\caption{Normalized IC-XT over time in MCF with various signalling sources}
\label{fig:lightsour}
\end{minipage}
\end{figure*}

\begin{figure*}[t]
\begin{minipage}[t]{\columnwidth}
\centering
	\includegraphics[width=0.99\columnwidth]{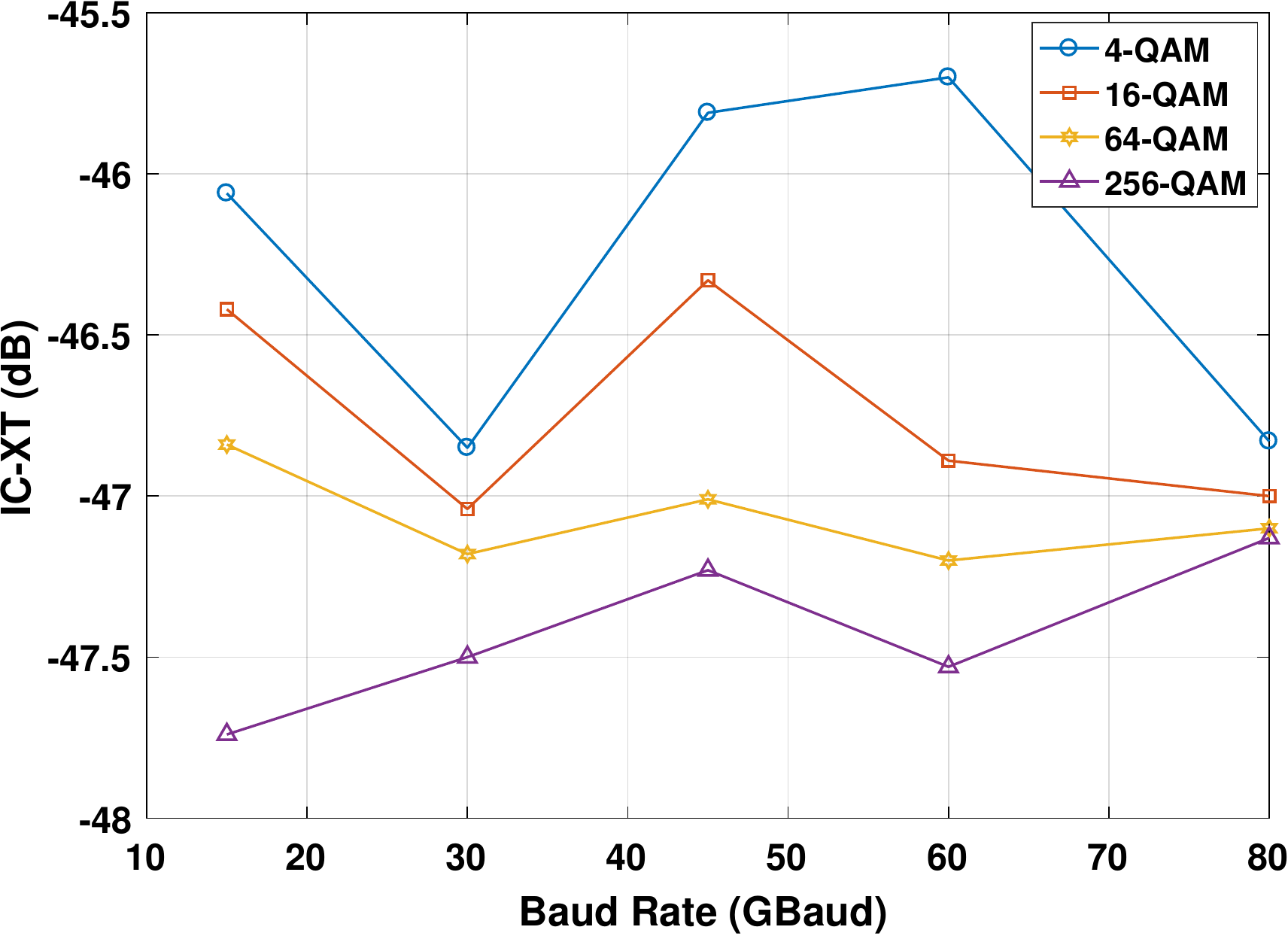}
	\caption{Static IC-XT for different baud rates and phase modulated formats}
	\label{fig:baudrate}
\end{minipage}%
\hfill 
\begin{minipage}[t]{\columnwidth}
\centering
	\includegraphics[width=0.99\columnwidth]{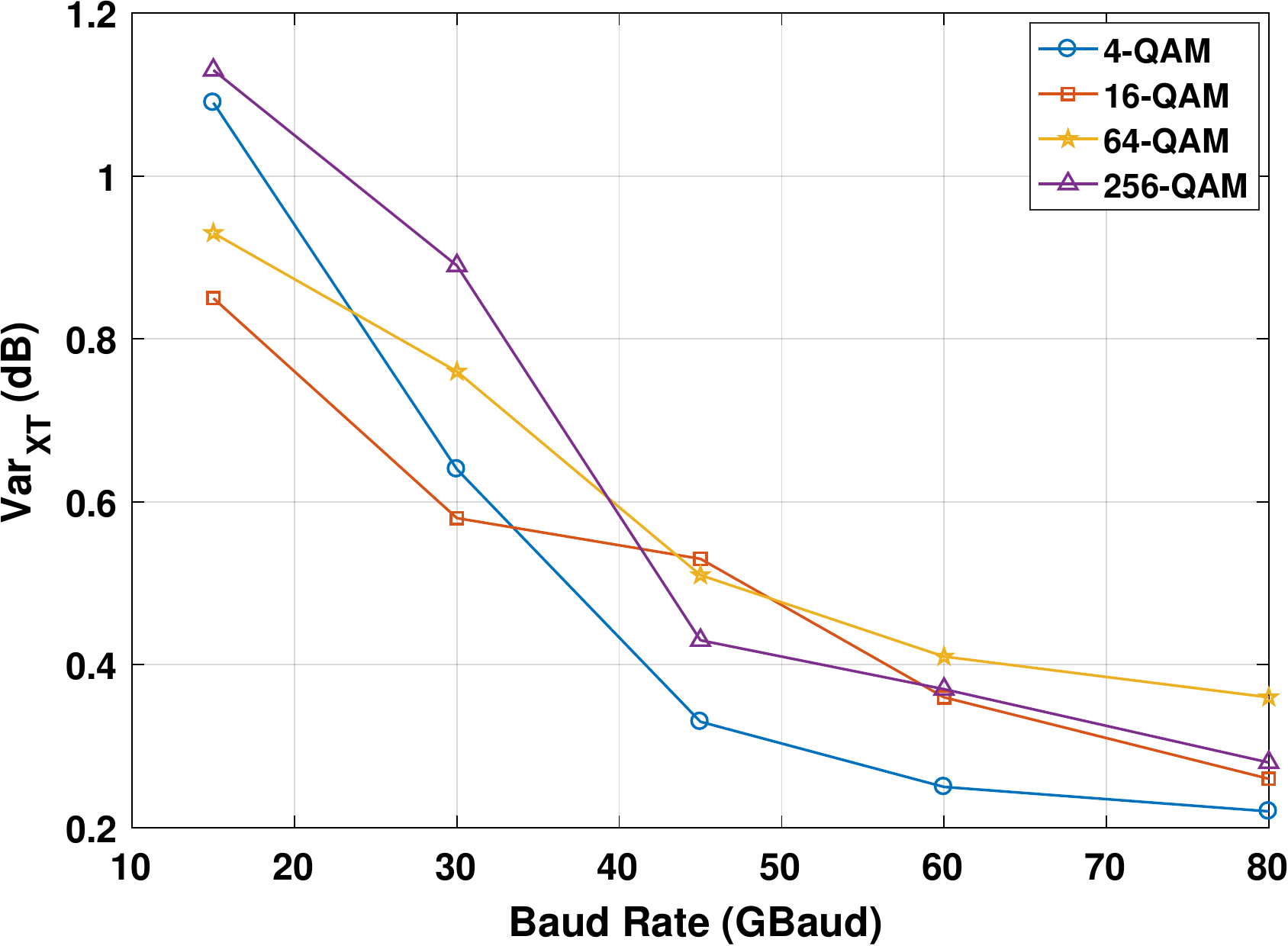}
	\caption{1-hour dynamic IC-XT  for different baud rates}
	\label{fig:baudrate_f}
\end{minipage}
\end{figure*}

\subsection{Static IC-XT Between Cores}
\label{subsec:pair}

First of all, we measured the static IC-XT between all the core pairs with an ASE light source operating at 1550 nm wavelength. The observed core-to-core (pairwise) IC-XT is presented as a bar chart in Fig.~\ref{fig:pairwise}, where each bar is an average result of an 1-hour measurement. For each subgroup, only one core was selected as the excited core (e.g. core 1), and then the IC-XT between the excited core and the other 7 cores (e.g. cores 2-8) were measured sequentially and individually. It can be easily seen that the pairwise IC-XT ranges from -71 dB to -44.85 dB roughly following the expectation, the smaller the core pitch between the core pair, the higher the IC-XT. To confirm the accuracy of the results, they are compared with the theoretical value calculated based on Eq.~(\ref{eq:meanXT}) (dashed lines). The comparison shows that the measured IC-XT fits the estimated value well when the core pitch between the cores is 35 $\mu$m, e.g. core 1 and core 3, core 4 and core 6. When the core pitch is 45~$\mu$m, e.g. core 5 and core 6, core 7 and core 8, the measured IC-XT is slightly higher than the calculated one. Based on the measurements without MCF, it was found that the IC-XT induced by the fan-in/out devices, ranging from -72 to -55~dB, contributed to these slight increases. However, when the core pitch is bigger than 55~$\mu$m, the observed IC-XT is much higher than the theoretical one. This is because the IC-XT from the fan-in/out devices dominates the measured results.  

According to the previous results, from subsection \ref{subsec:LSaBR} to subsection \ref{subsec:tem}, we measured the IC-XT between core 1 and core 3, where core 1 was the excited core. Since the core pitch between them was the smallest (35~$\mu$m), the highest (worst case) IC-XT should be obtained. While for subsection \ref{subsec:ncore}, multiple cores were excited, the setup will be described in detail in the related subsection.

\subsection{Signalling Source and Baud Rate}
\label{subsec:LSaBR}


\begin{figure*}[t]
\begin{minipage}[t]{\columnwidth}
\centering
	\includegraphics[width=0.99\columnwidth]{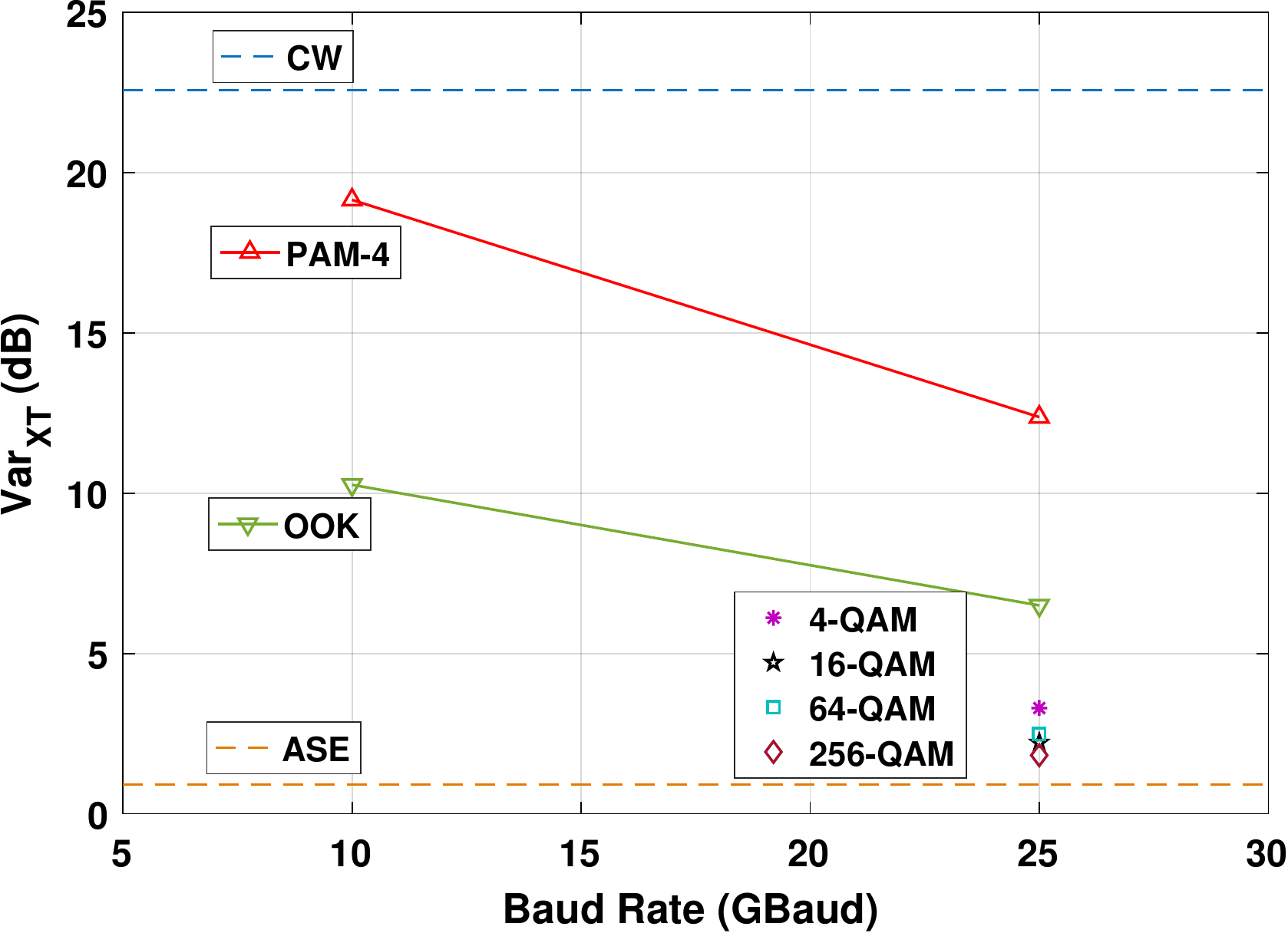}
	\caption{12-hours dynamic IC-XT and IC-XT PDF for various signals}
	\label{fig:varxt}
\end{minipage}%
\hfill 
\begin{minipage}[t]{\columnwidth}
\centering
	\includegraphics[width=0.99\columnwidth]{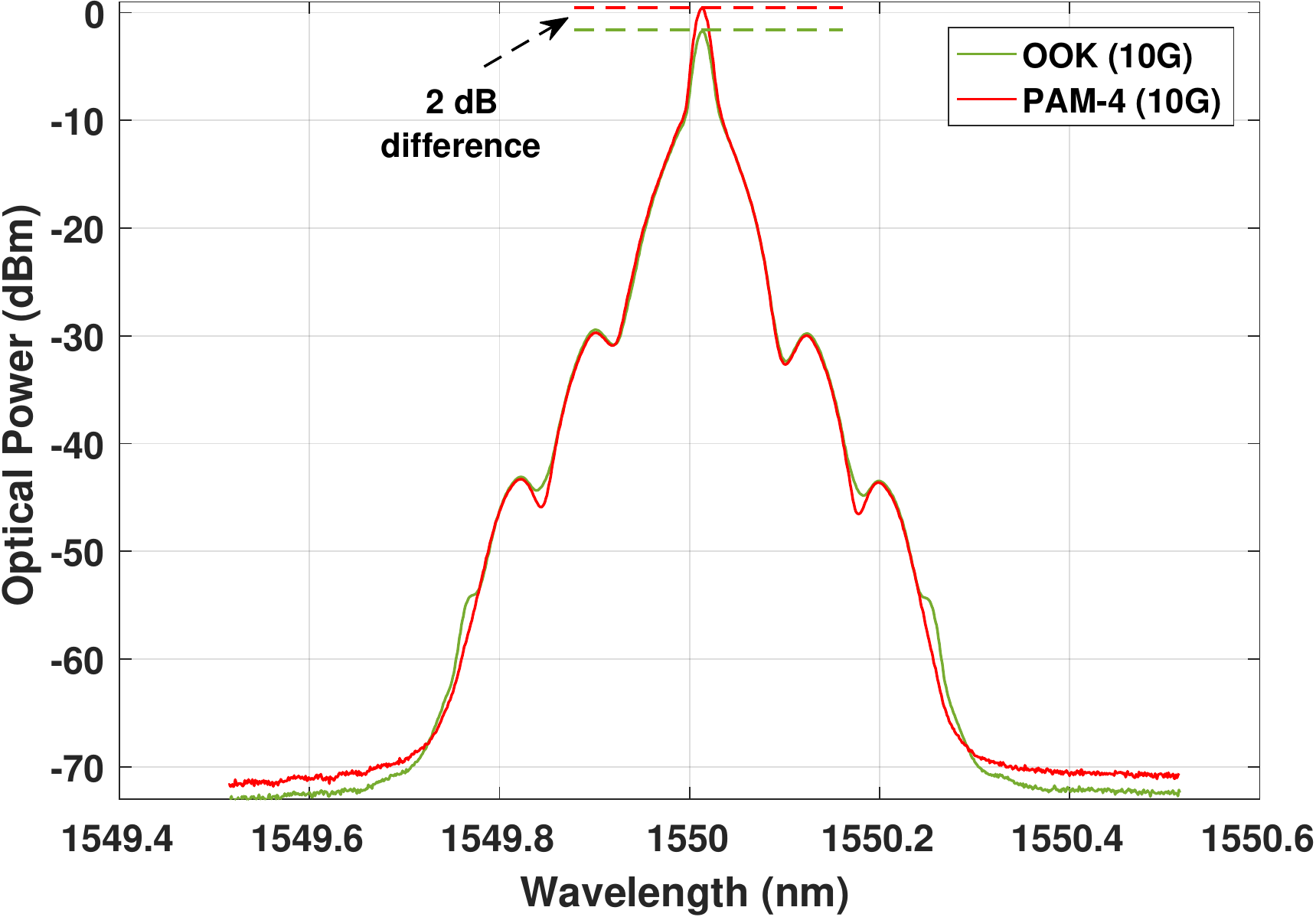}
	\caption{Optical Spectrum of OOK and PAM-4 (Resolution 0.02 nm)}
	\label{fig:sp}
\end{minipage}
\end{figure*}

Up to now, CW light signals, ASE signals or intensity modulated signals, i.e. OOK, have been widely used to stimulate and study IC-XT \cite{Rademacher_2017,Hayashi_2011b,Alves_2018}. Figure~\ref{fig:lightsour} depicts the observed normalized IC-XT in the TA-MCF induced by these three signalling sources over 12 hours, where each point was the short term average IC-XT over 25~ms. It can be clearly seen that for CW and OOK signals, the induced IC-XT changes rapidly over time. These vast fluctuations can be attributed to macro bending, twists and structural fluctuations on the fiber \cite{Hayashi_2011a}. Moreover, as seen, these fluctuations can be affected by the type of source signals and baud rate. Among the various cases analyzed in Fig.~\ref{fig:lightsour}, the IC-XT measured using a narrow CW laser source has the highest level of dynamicity, which extends up to 22.58 dB. On the contrary, IC-XT measured using the broadband (150 GHz) ASE light source is stable at its statistical mean and it achieves the smallest dynamic range of 0.916 dB. Note that, the MCF utilized in this work is much shorter than the one adopted in \cite{Rademacher_2017} as it is designed for data center usage, also, the fiber type, i.e. trench-assisted rather than step-index, layout and core pitch are different. All these parameters can significantly affect the IC-XT behavior, and thus, compared to the result shown in \cite{Rademacher_2017}, IC-XT with higher dynamicity (i.e. 0.916, 6.58 and 4 dB higher dynamic IC-XT for ASE, CW and 10G OOK sources, respectively) was observed. In contrast, IC-XT results from the intensity modulated OOK signals achieve the second lowest variation. Whereas, a 15 GBaud increase in the baud rate can further reduce the dynamicity of IC-XT by an extra 5 dB. Furthermore, by analyzing the observed IC-XT for ASE and OOK (25~G) sources, it is found that the average IC-XT of every 1-hour measurement (i.e. -46.34 to -45.5 dB and -46.08 to -45.8 dB, respectively) was within $\pm$0.5 dB to the average IC-XT over 12 hours (i.e. -45.95 dB and -45.92 dB, respectively). In addition, in dynamic DCNs, attributing to the fast circuit and packet switching, the links may only serve the signals for a few nanoseconds, millisecond, seconds, minutes or hours. Therefore, most of the following IC-XT results were obtained over a time window of 1 hour to study the IC-XT behavior in this short period of time. The investigation of the time window effect on the observed IC-XT will be presented in Section~\ref{accu}.

Apart from the previous three types of signalling sources discussed, the behavior of IC-XT induced by I-Q modulated signals was also investigated. Figure~\ref{fig:baudrate} depicts the observed static IC-XT in TA-MCF generated by four types of QAM signals with baud rate ranging from 15 to 80 GBaud. Each point on the graph has a 1-hour measurement time. As it can be seen, with the increase in the order of QAM format, the observed static IC-XT decreases and the maximal difference between the observed IC-XT induced by 4-QAM and 256-QAM signals for a given baud rate is 1.8 dB. Moreover, it was found that when the QAM order increased from 4 to 256, the OSNR of the signals reduced from 48.6 dB to 46.3~dB. Therefore, it is believed that these observation results are due to the OSNR variation with QAM order rather than that the IC-XT was intrinsically affected. Since OSNR difference can be observed on any transmission and network experiments, the figure points out that similar observations should not be wrongly perceived as the change from IC-XT itself. In addition, Fig.~\ref{fig:baudrate} clearly shows that the static IC-XT is relatively stable (slight fluctuations may result from the 1-hour time window) when the baud rate increases from 15 to 80 GBaud. That is to say, the static IC-XT is baud rate independent. However, the dynamic IC-XT ($Var_{XT}$), as shown in Fig.~\ref{fig:baudrate_f}, is inversely proportional to the baud rate as explained in Section \ref{Dynamic}. To some extent, it extends the experimental results in \cite{Rademacher_2017} and confirms that the conclusion from \cite{Rademacher_2017} on baud rate effect is also suitable for cases using different orders of I-Q modulated signals.


To summarize this subsection, the IC-XT variations for all the investigated light sources and modulation formats have been presented in Fig.~\ref{fig:varxt}. Each point on the figure is an average result of a 12-hour measurement, and all the results were obtained with 1550~nm wavelength. According to \cite{Hayashi_2014}, the wider the signal bandwidth, the lower the IC-XT variation. Therefore, as illustrated in Fig.~\ref{fig:lightsour}, IC-XT induced by narrow CW source and broadband ASE light set the upper and lower bounds, respectively. In terms of intensity modulated formats, the strong optical carriers and/or carrier-to-signal power ratios lead to a stronger interference at PMPs and thus, high stochastic variations in IC-XT intensity can be observed. In addition, IC-XT for PAM-4 signals provides higher variations compared to that of OOK signals over all studied baud rates. This can be explained by the fact that compared to OOK signalling more non-zero intensity levels exist in PAM-4, which leads to more than 30\% increase (by calculation) in the optical carrier-to-signal ratio as result of optical light being present more often. It increases the probability of attaining a non-zero level over PMPs across the MCF. Experimentally, from the optical spectrum of the OOK and PAM-4 signals shown in Fig.~\ref{fig:sp}, a 2~dB difference in optical carrier-to-signal power ratio is observed.

In contrast, IC-XT for QAM signals fluctuates much less than that of the OOK and PAM-4 signals and all QAM cases achieve a variation level close to that of the ASE scheme. This is owe to the removal of the optical carrier. Among the four QAM formats, 4-QAM exhibits the highest variation whilst 256-QAM provides the lowest. The $>$5~dB reduction  in dynamic IC-XT, as a result of a 15 GBaud baud rate increase, is also evident for the two types of intensity modulated signals. However, as the measurement times for QAM formats over the remaining baud rates were 1 hour rather than 12 hours, results in Fig.~\ref{fig:baudrate_f} are not included in Fig.~\ref{fig:varxt}. Nevertheless, the 1-hour observation of IC-XT presented in Fig.~\ref{fig:baudrate_f} shows that for every QAM format, a more than 1 dB reduction in dynamic IC-XT can be achieved by increasing the baud rate from 15 to 80 GBaud.

\subsection{Temperature}
\label{subsec:tem}

\begin{figure*}[t]
\begin{minipage}[t]{\columnwidth}
\centering
	\includegraphics[width=0.99\columnwidth]{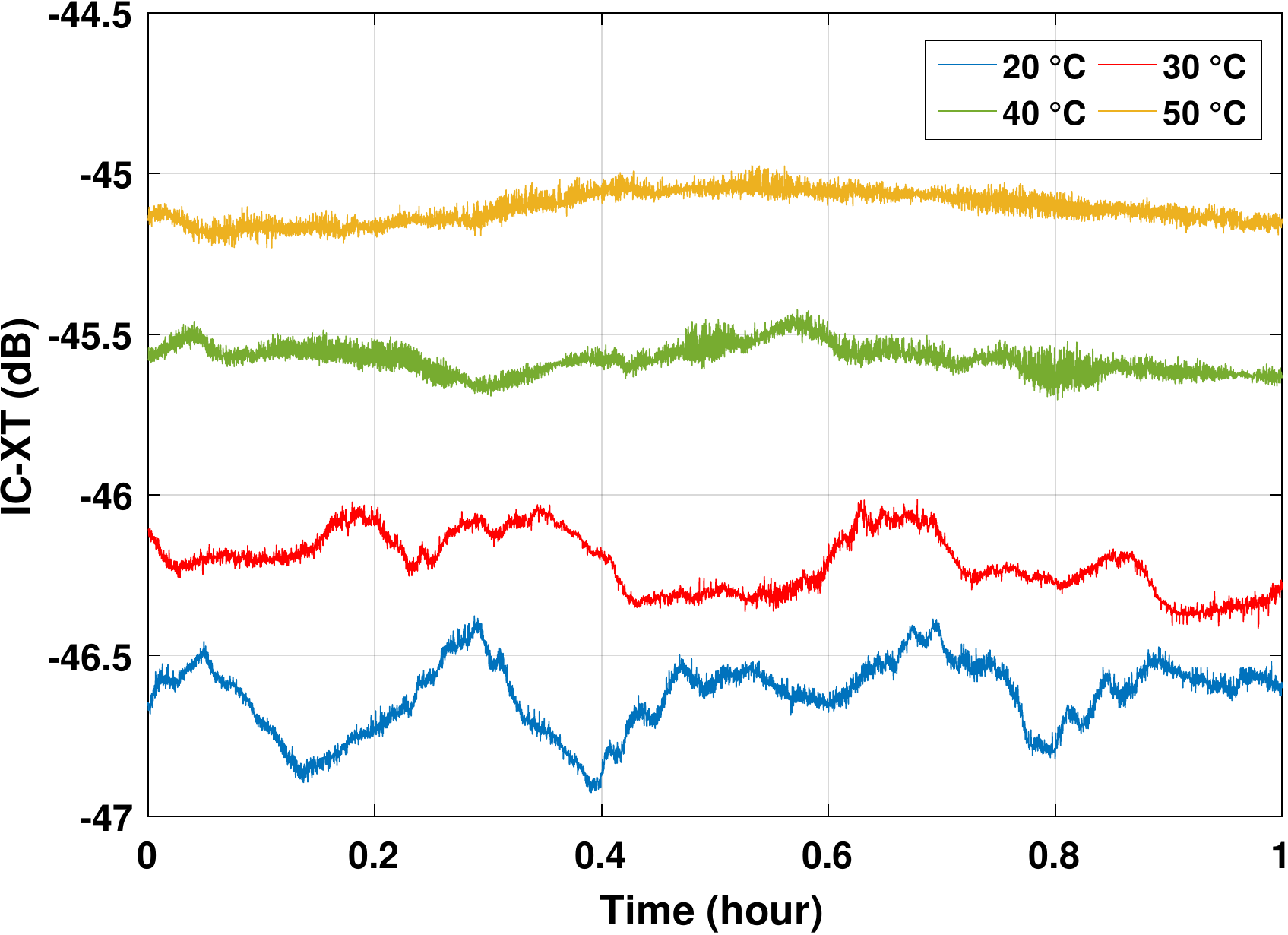}
	\caption{Effect of temperature on time-dependent IC-XT (ASE)}
	\label{fig:temXT}
\end{minipage}%
\hfill 
\begin{minipage}[t]{\columnwidth}
\centering
	\includegraphics[width=0.99\columnwidth]{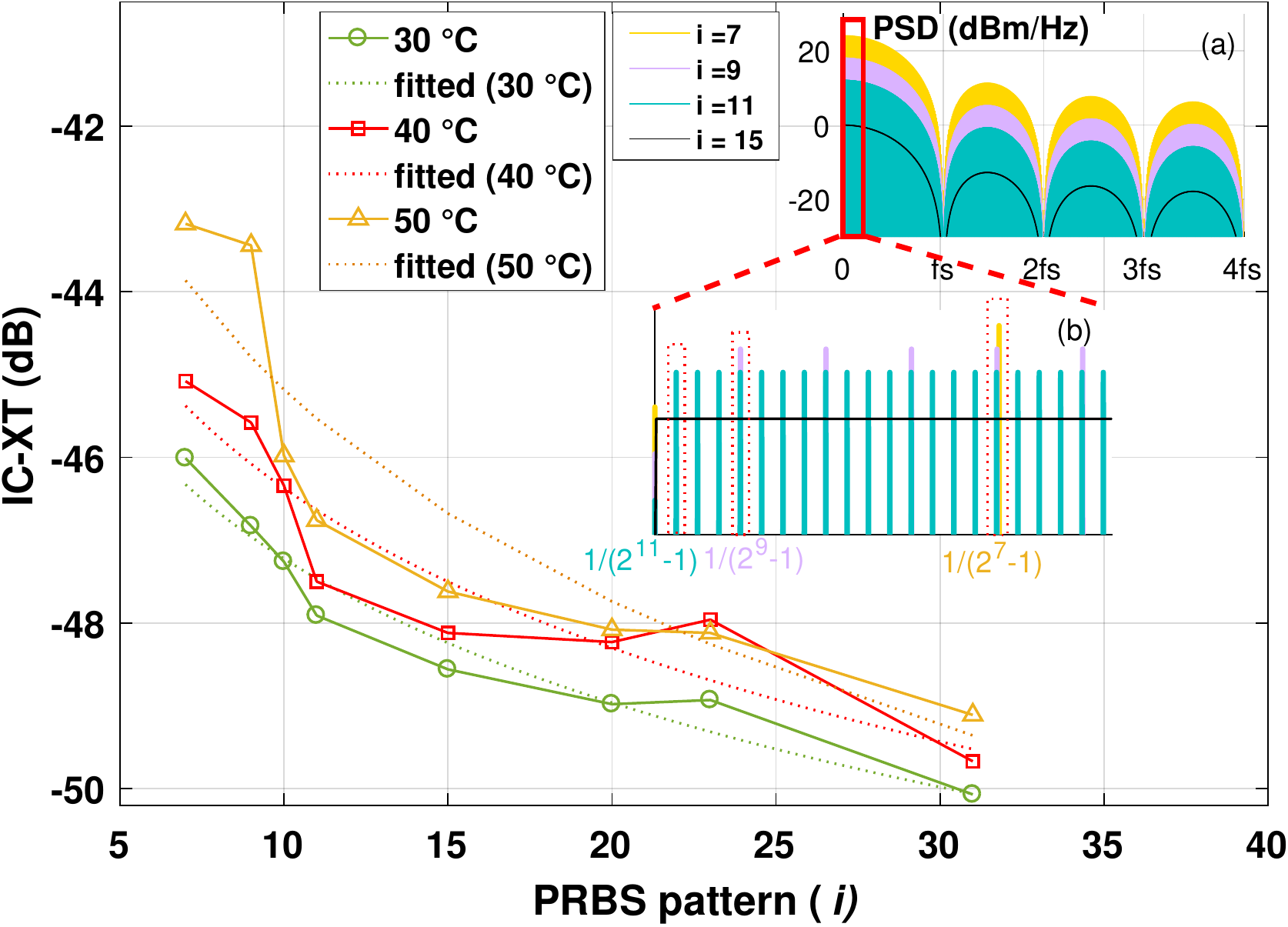}
	\caption{Effect of PRBS length on static IC-XT (25G-OOK)}
	\label{fig:temprbs}
\end{minipage}
\end{figure*}

\begin{figure*}[t]
\begin{minipage}[t]{\columnwidth}
\centering
\includegraphics[width=0.99\columnwidth]{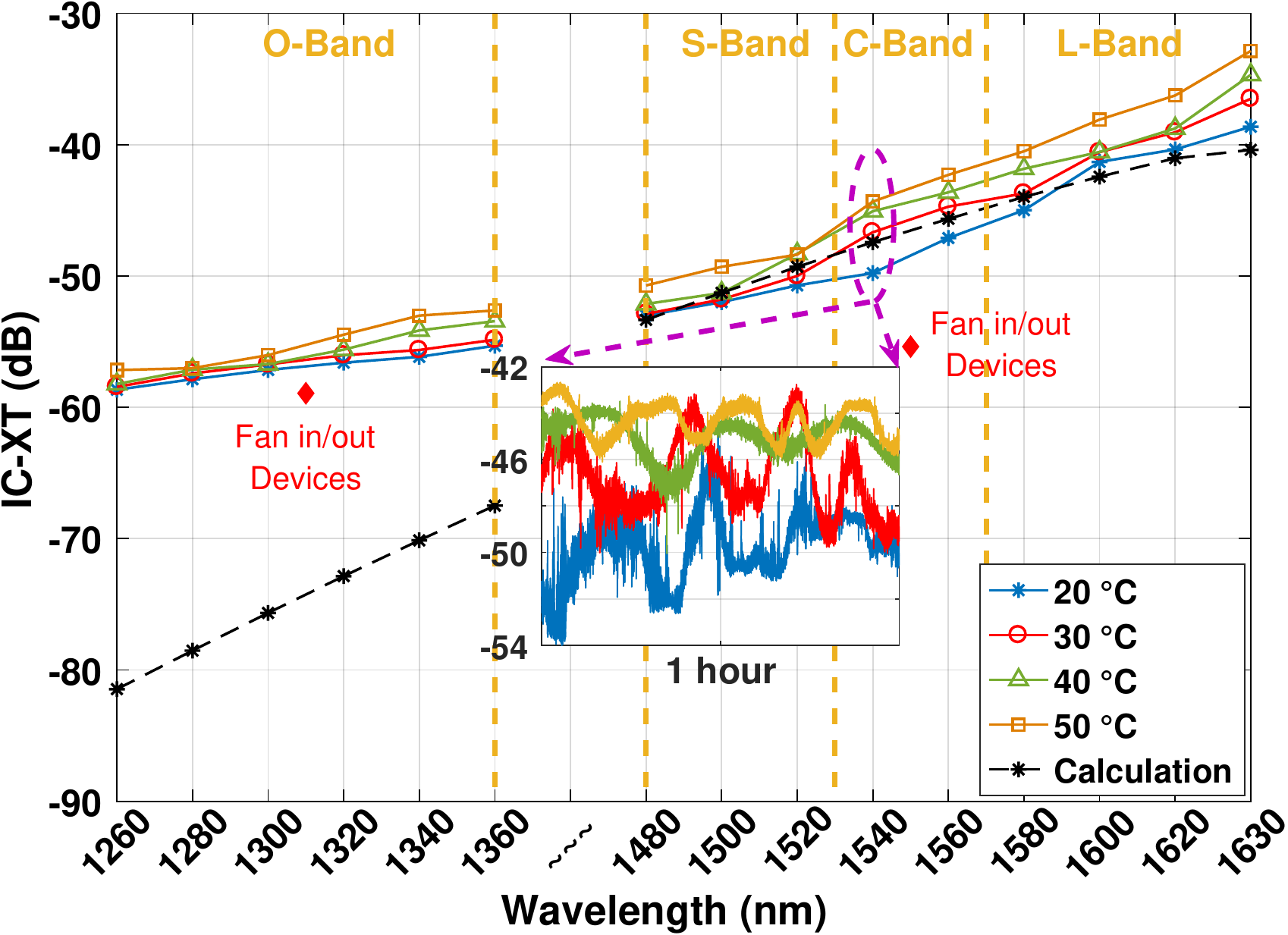}
\caption{Effect of temperature and wavelength on static IC-XT (25G-OOK)}
\label{fig:temwav}
\end{minipage}%
\hfill 
\begin{minipage}[t]{\columnwidth}
\centering
\includegraphics[width=0.99\columnwidth]{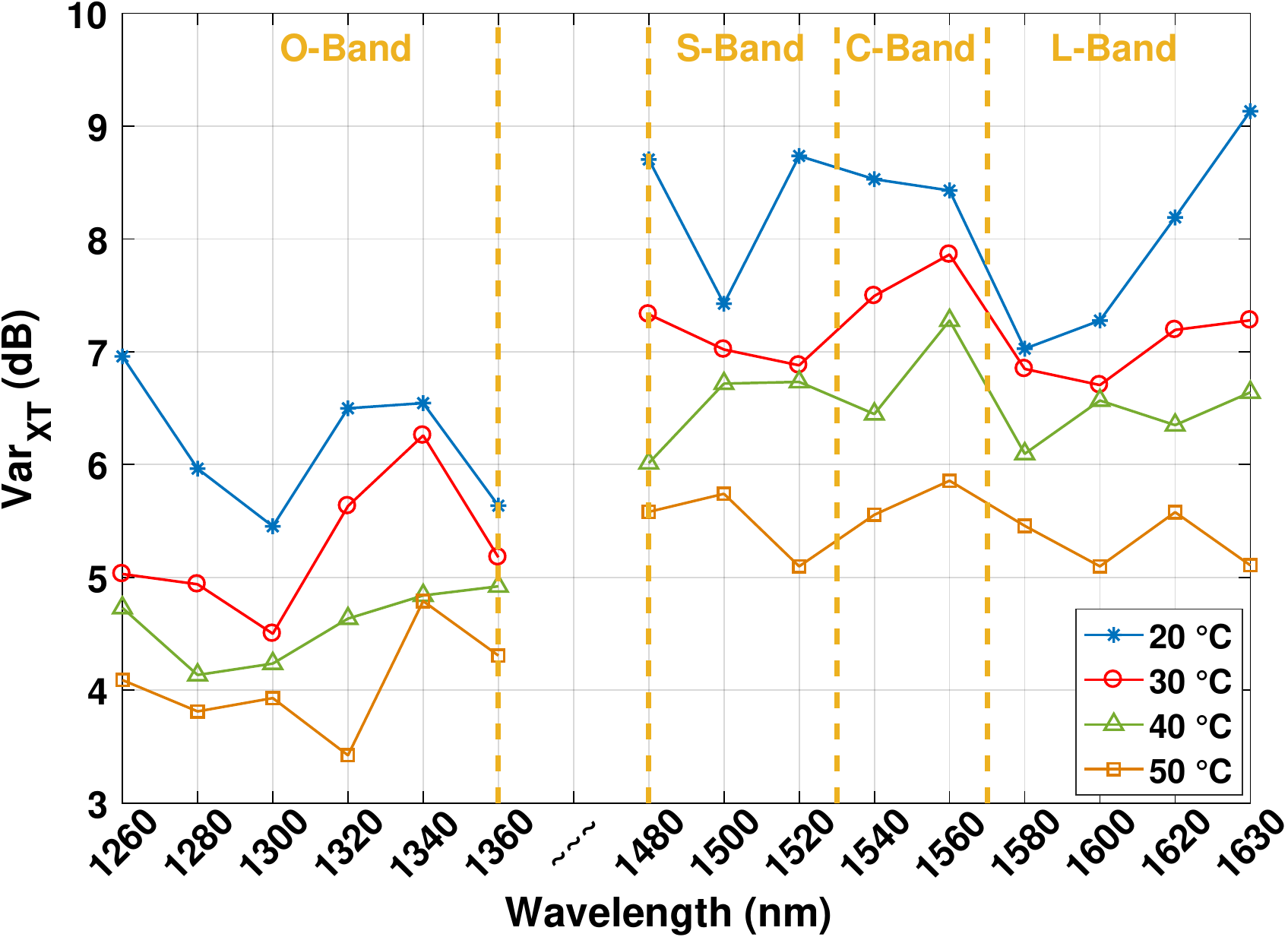}
\caption{Effect of temperature on dynamic IC-XT (25G-OOK)}
\label{fig:temwavf}
\end{minipage}
\end{figure*}

Based on Eqs.~(\ref{eq:meanXT}) and (\ref{eq:ka}), it can be deduced that considerable IC-XT increase would be achieved, even when the refractive index contrast between the core and the cladding reduces slightly. For instance, based on calculation, a $5\times10^{-5}$ variation in refractive index can lead to {$>$}1 dB static IC-XT increase between core 1 and core 3 over the utilized TA-MCF. Figure~\ref{fig:temXT} depicts the observed time-dependent IC-XT in the TA-MCF for temperature ranging from 20 to 50 $^\circ$C over 1-hour measurement. Since ASE signal can lead to the lowest IC-XT variation which has been shown in Fig.~\ref{fig:varxt}, the 1550~nm ASE signalling source was adopted to explore the effect from temperature on IC-XT. It can be seen that a 30~$^\circ$C temperature increase leads to a 1.5~dB increase in static IC-XT and 0.3~dB reduction in dynamic IC-XT, which can be translated into a dependence coefficient of 0.05 dB/K and -0.01~dB/K, respectively. According to \cite{Yang_2014,Wang_2014, Ismail_2009}, refractive index changes with temperature change. Moreover, the changes on the refractive indexes of core, cladding and trench may be different since they have different properties, including radius, original index and so on \cite{Ismail_2009}. These differences may change the index contrasts which contribute to static IC-XT increase. While the dynamic IC-XT reduction is due to the increase in skew \cite{Rademacher_2017}, as explained in Section \ref{Dynamic}. Similar trends are observed in Figs.~\ref{fig:temprbs} and \ref{fig:temwav}, where the realistic 25~GBaud OOK signals were utilized as the signalling sources and each point on the graphs was measured over a 1-hour period.

\subsubsection{PRBS Length}
\label{subsubsec:PRBS}

Figure~\ref{fig:temprbs} shows the effect of PRBS length on the static IC-XT. Each point on the figure is the 1-hour static IC-XT considering 144,000 samples. It can be seen in Fig.~\ref{fig:temprbs}, an increase in the PRBS length from 2$^{7}$-1 to 2$^{31}$-1 contributes to in exceed of 4~dB decrease (i.e. from -46.01 to -50.07 dB) in static IC-XT at 30~$^\circ$C and 6~dB (i.e. from -43.18 to -49.11 dB) at 50~$^\circ$C. In particular, it is found that the relationship between crosstalk and PRBS length for this type of MCF can roughly fit the function, $crosstalk=a \text{log}_2i+b$, where the coefficients a and b under temperatures of 30~$^\circ$C, 40~$^\circ$C and 50~$^\circ$C equal to -1.7 and -41.4, -1.9 and -40.0, -2.6 and -36.7, separately. As seen, some of the points in the figure do not follow the fitting line, e.g. when temperature was 30~$^\circ$C and PRBS pattern was 23, temperature was 40~$^\circ$C and PRBS pattern was 23. Both of them were higher than the trend line, we believe that this correlation for different temperatures between experimental deviations from the fitting can be explained by the chronological order of the measurements, for each PRBS length the fiber was subjected to one heat cycle. To explain the correlation between the static IC-XT and the PRBS length, the PSD of signals with various PRBS patterns has been presented in inset (a) and its zoomed-in figure (particularly for the part inside the red rectangle), inset (b), where PRBS15 was the benchmark and all the other patterns (PRBS7, 9 and 11) were repeated to make all sequences the same length with it. It can be seen that, with the increase of the PRBS order, the frequency components for a given range increases while the power of each component (inset(a)) reduces to maintain the same total power for each case. The spacing between the two continuous components (inset(b)) also reduces as it is given by the bit rate divided by the length of the PRBS [i.e. $1/(2^i-1)$]~\cite{Wiley}. Thus, the reasons for the correlation are: (i) an increase in the randomness in the transmission pattern, reducing the coherent interference and, (ii) an increase in the spectral content in the frequency domain as a result of higher spectral sampling.

\subsubsection{Wavelength}
\label{subsubsec:wav}

\begin{figure}[t]
	\centering
	\includegraphics[width=0.99\columnwidth]{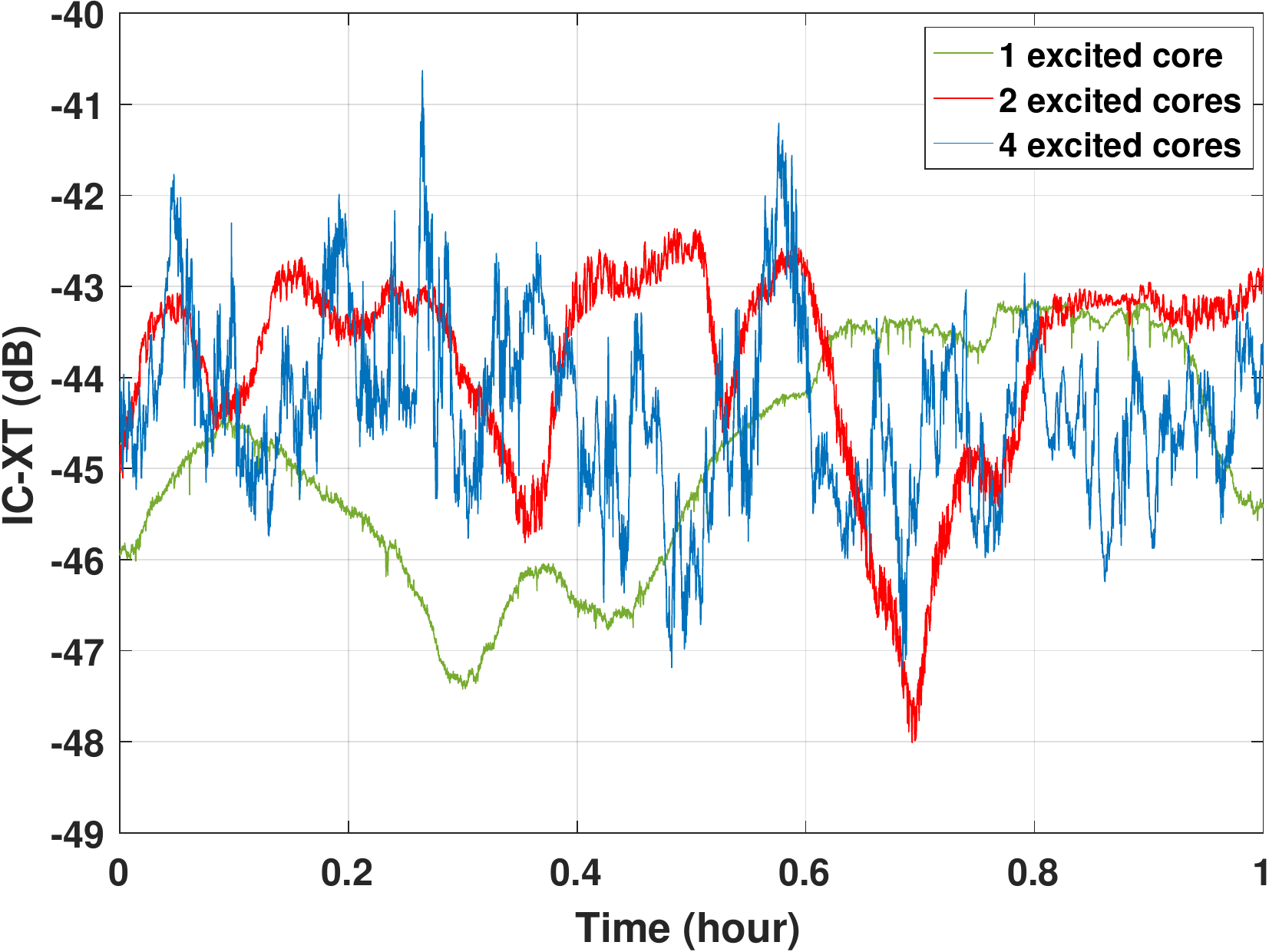}
	\caption{IC-XT over time for different numbers of excited cores}
	\label{fig:multicores}
\end{figure}

The effects of temperature and operational wavelength on the static IC-XT and dynamic IC-XT can be observed from Fig.~\ref{fig:temwav} and Fig.~\ref{fig:temwavf}, respectively. As it can be seen in Fig.~\ref{fig:temwav}, as the operational wavelength increases from 1480~nm (S-band) to 1630~nm (L-band), the static IC-XT increases by around 17~dB for every temperature studied, which can be translated into crosstalk-wavelength dependence coefficient of 0.113 dB/nm. The reason for this dependence of IC-XT on wavelength, which has been mentioned in Section~\ref{subsec:Sta}, is that the mode field diameter enhances with longer wavelengths, leading to wider mode field area overlapping between the adjacent cores and in turn, higher power leakage between the cores. Furthermore, it is observed that as the operational wavelength moves from O-band to S-, C- and L-bands, the enhancement of IC-XT as a result of temperature increase also increases by over 100\%, averaging from 0.06 dB/K to 0.13 dB/K. This is due to the fact that the observed IC-XT for O-band signals after 1 km of MCF transmission ({$\approx$} -81.5 to -67.5~dB, calculated based on Eqs.~(\ref{eq:meanXT}) and (\ref{eq:ka})) is dominated by the IC-XT generated inside the fan-in/out devices ({$\approx$} -58~dB at 1310~nm, measured). Moreover, it should be mentioned that, as the sensitivity of the coupling coefficient of the adopted TA-MCF to the wavelength change varies in different bands, the slope of the indicator line (black line) for O-band is bigger than that of the S-, C-, L-bands. In addition, from the zoomed-in figure in Fig.~\ref{fig:temwav}, it can be observed that, the fluctuation range of IC-XT (dynamic IC-XT) changes with the temperature increase.

To better understand the relationship between the dynamic IC-XT with wavelength and temperature, for the realistic transmission scenario, Fig.~\ref{fig:temwavf} is shown. As the O-band IC-XT dominantly came from the fan-in/out devices, our study focus on the S-, C- and L-bands IC-XT variation. It is observed that a 30~$^\circ$C increase in temperature leads to a 2.6~dB IC-XT variation reduction, which means that the temperature-crosstalk variation dependence coefficient is -0.09~dB/K. The absolute value is 0.08 dB/K higher than that of the ASE scenario showcased in Fig.~\ref{fig:temXT}. In addition, the graph also indicates that the dynamic IC-XT is wavelength independent as it is relatively stable when the operational wavelength goes up from S-band to L-band.

\subsection{Number of Excited Cores}
\label{subsec:ncore}

All the previous investigations concentrate on the behaviour of IC-XT results from one excited core adjacent to the target core. To evaluate the behaviour if multiple cores are stimulated simultaneously, more than one core were excited with the same signalling source using a 1x4 splitter as shown in Fig.~\ref{fig:setup}. The observed results are presented in Fig.~\ref{fig:multicores}, where core 5 was the target core and core 3 was the excited core for a singular excited core case; cores 3 and 7 were the excited cores for dual excited core case; cores 3, 6, 7 and 8 were the excited cores for quadruple excited core case. It can be easily seen that, the IC-XT fluctuates slowly when there is only one excited core, however, when the number of excited core increases, the fluctuation speed considerably increases. This can be explained by the fact that the number of the PMPs rises with the increase of the number of excited cores. Although, based on the previous results, no distinct changes were observed in the static or dynamic IC-XT when the fluctuation speed changed, the speed of the IC-XT fluctuations should be characterized for MCF applications in the real world. Because it may affect the design of crosstalk-tolerant adaptive techniques and indicates how often the service will be interrupted \cite{Alves_2017}. Specifically, by estimating the IC-XT fluctuation speed through analyzing the number of peaks and bottoms on the lines, it is found that the IC-XT fluctuation speeds for 2 and 4 excited cores cases are 7 and 30 times of that of the 1 excited core case, respectively. That is to say, IC-XT fluctuation speed roughly increases by a factor of 7.4 per core. It should be noted that, the observed result in this section can greatly fit the result in \cite{Puttnam_2018} even the fibers and light sources used were completely different in terms of fiber length, core pitch and core layout, which indicates that the excited core number has higher priority on affecting the speed of IC-XT fluctuation than them.

\section{Effect of Measurement Method on Observed IC-XT and Analysis of IC-XT Step Distribution}
\label{accu}
The transmission parameters mentioned in the previous section can internally affect the IC-XT behavior. In the IC-XT measurement process, the observed IC-XT properties can also be affected by the measurement time \cite{Alves_2019} and the averaging time of the power meter or the interval time between the samples. In this section, the effects of time window and averaging time on the value of the observed IC-XT induced by different light sources are investigated. Moreover, an analysis on the IC-XT step distribution is presented.

\begin{figure*}[b]
\begin{minipage}[t]{\columnwidth}
\centering
\includegraphics[width=0.99\columnwidth]{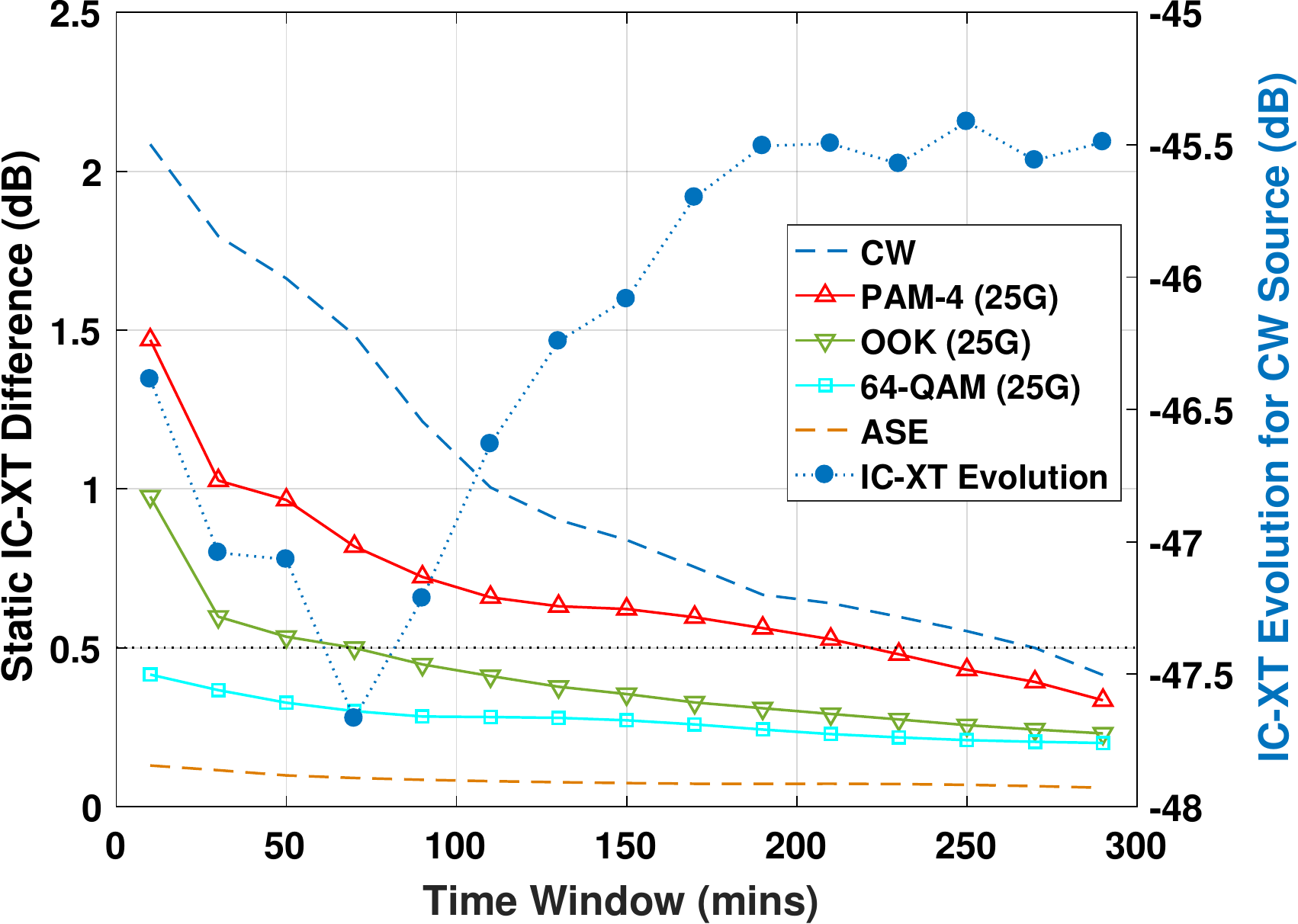}
\caption{Effect of time window on static IC-XT for various signals}
\label{fig:winxt}
\end{minipage}%
\hfill 
\begin{minipage}[t]{\columnwidth}
\centering
\includegraphics[width=0.99\columnwidth]{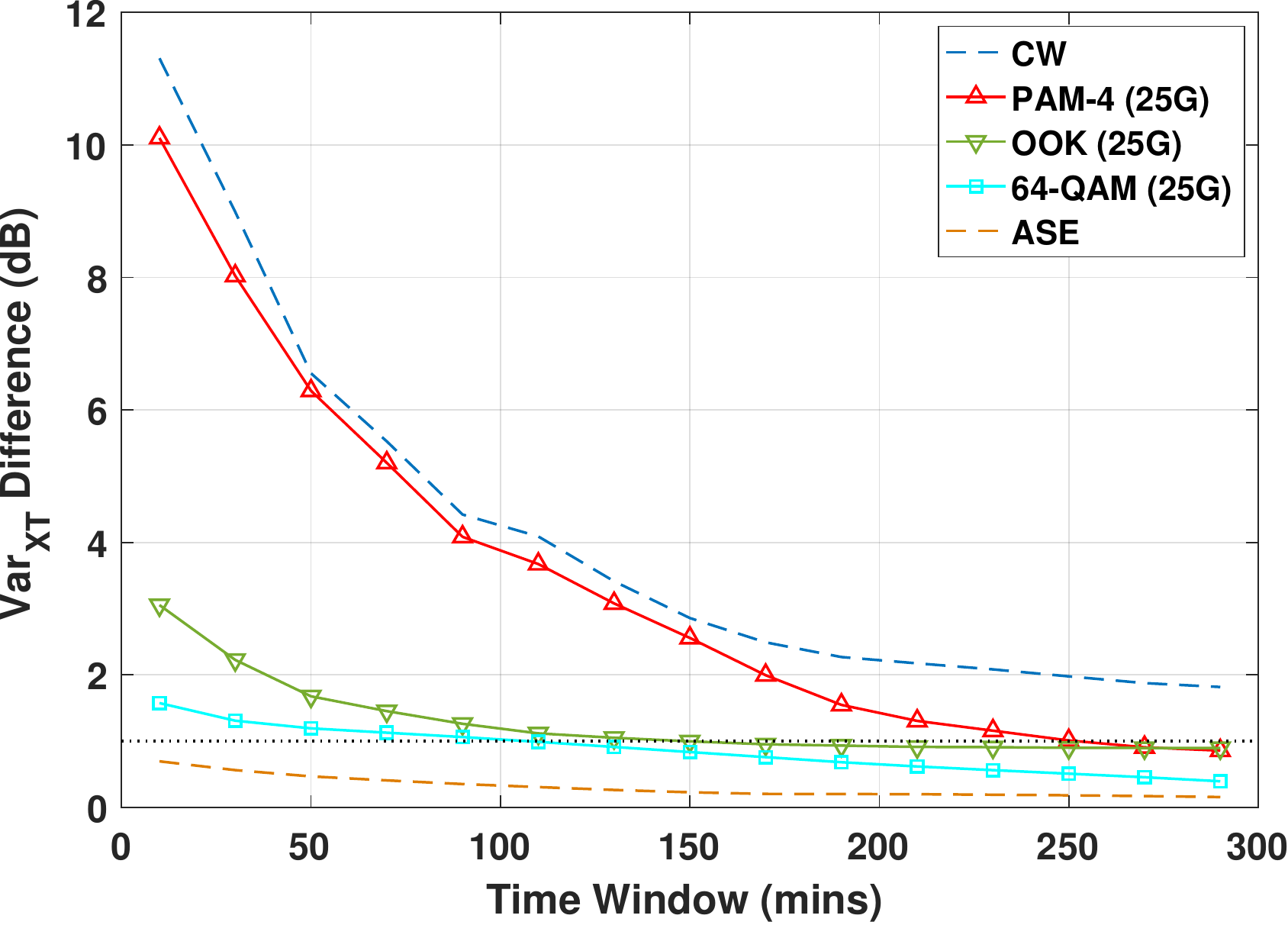}
\caption{Effect of time window on dynamic IC-XT for various signals}
\label{fig:winvar}
\end{minipage}
\end{figure*}
 
\subsection{Time Window}
\label{subsec:TW}
Figures~\ref{fig:winxt} and \ref{fig:winvar} show the effect of the measurement time window on the static and dynamic IC-XT induced by difference signalling sources, where the window of 12 hours is the benchmark since the convergence value for the 12 hours' data is 97\% (with 2\% oscillation in 10-12 hours) compared to that of the 30 hours' experimental data ($>$90\% convergence is considered as an accurate representation of the parent population, therefore is adequate for statistical analysis~\cite{Coleman_2009}), which shows a steady behavior with oscillation of 0.3\% in the last two hours (28-30 hours). For the intensity modulated signals, PRBS 15 were utilized. It can be seen that, the degree of the effect from time window on both the static and dynamic IC-XT changes with signalling source change, among which the time window has higher degree of effect on the dynamic IC-XT (i.e. up to 11.31 dB difference from the benchmark) than that on the static IC-XT (i.e. up to 2.09 dB difference from the benchmark). By combing the information in Fig.~\ref{fig:varxt}, which shows the dynamicity of IC-XT induced by different sources, and these two figures, it can be found that the absolute value of difference increases with the stability of the IC-XT decrease. For example, for ASE source, even the time window was 10 mins, only 0.13 dB difference compared to the 12-hours' result on static IC-XT and 0.68 difference on dynamic IC-XT were observed. In contrast, for CW source, although the time window went up to 300 mins, the differences were around 0.4 dB and 2 dB, respectively. In addition, as an example, the evolution of IC-XT for CW source is also shown in Fig.~\ref{fig:winxt}. These two figures provide the information of how long the measurement should be for different static and dynamic IC-XT requirements. For instance in this paper, the 1-hour measurements with OOK and QAM sources might lead to 0.53~dB and 0.33~dB averaged differences from the benchmark on the static IC-XT, respectively.

\subsection{Averaging Time}
\label{subsec:Avg}

\begin{figure}[t]
	\centering
	\includegraphics[width=0.99\columnwidth]{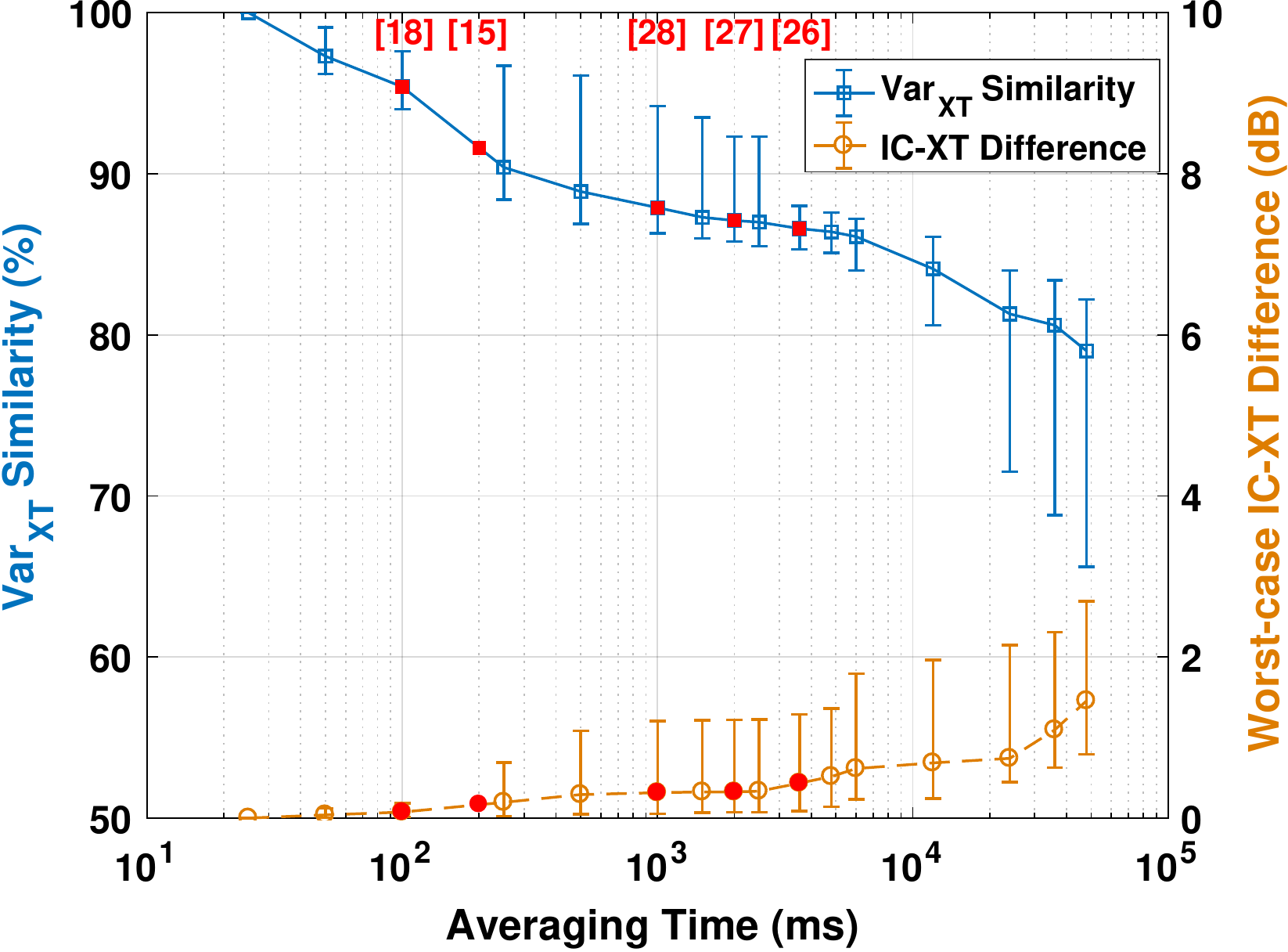}
	\caption{Effect of averaging time on IC-XT measurement}
	\label{fig:avg}
\end{figure}

Various averaging times or interval times between samples have been considered in the existing researches on IC-XT evaluation, including 100~ms \cite{Luís_2016}, 200~ms \cite{Rademacher_2017}, $<$1~s \cite{Alves_2019}, 2~s \cite{Alves_2018} and 3.6~s \cite{Luís_2015}. Different averaging time may change the observed IC-XT properties in varying degrees. To explore the effect of averaging time on dynamic IC-XT and the worst-case IC-XT, which are significant for MCF application in practice, Fig.~\ref{fig:avg} is shown. Static IC-XT is not considered in this figure since no distinct changes had been observed when the averaging time changed. To obtain the IC-XT for different averaging time, we first measured the IC-XT with the smallest averaging time (25 ms) and then, re-sampled the data with different rates and calculated the short-term average IC-XT accordingly. As seen, when the averaging time changes from 25~ms to 48~s, more than 20\% decrease in the similarity of dynamic IC-XT and nearly 2~dB difference in the worst-case IC-XT can be observed (25 ms is the benchmark). It should be noted that, for different light sources, the effects are different, and thus, the error-bar for each point is also provided. The figure also indicates that to achieve 90\% similarity to the benchmark in dynamic IC-XT, the averaging time should be smaller than 250 ms, at which point the difference in the worst-case IC-XT is 0.2 dB. In addition, compared with aforementioned researches, the results in this paper achieve 5\% - 13\% higher similarity to the benchmark on dynamic IC-XT.

\begin{figure*}[b]
\begin{minipage}[t]{\columnwidth}
\centering
	\includegraphics[width=0.99\columnwidth]{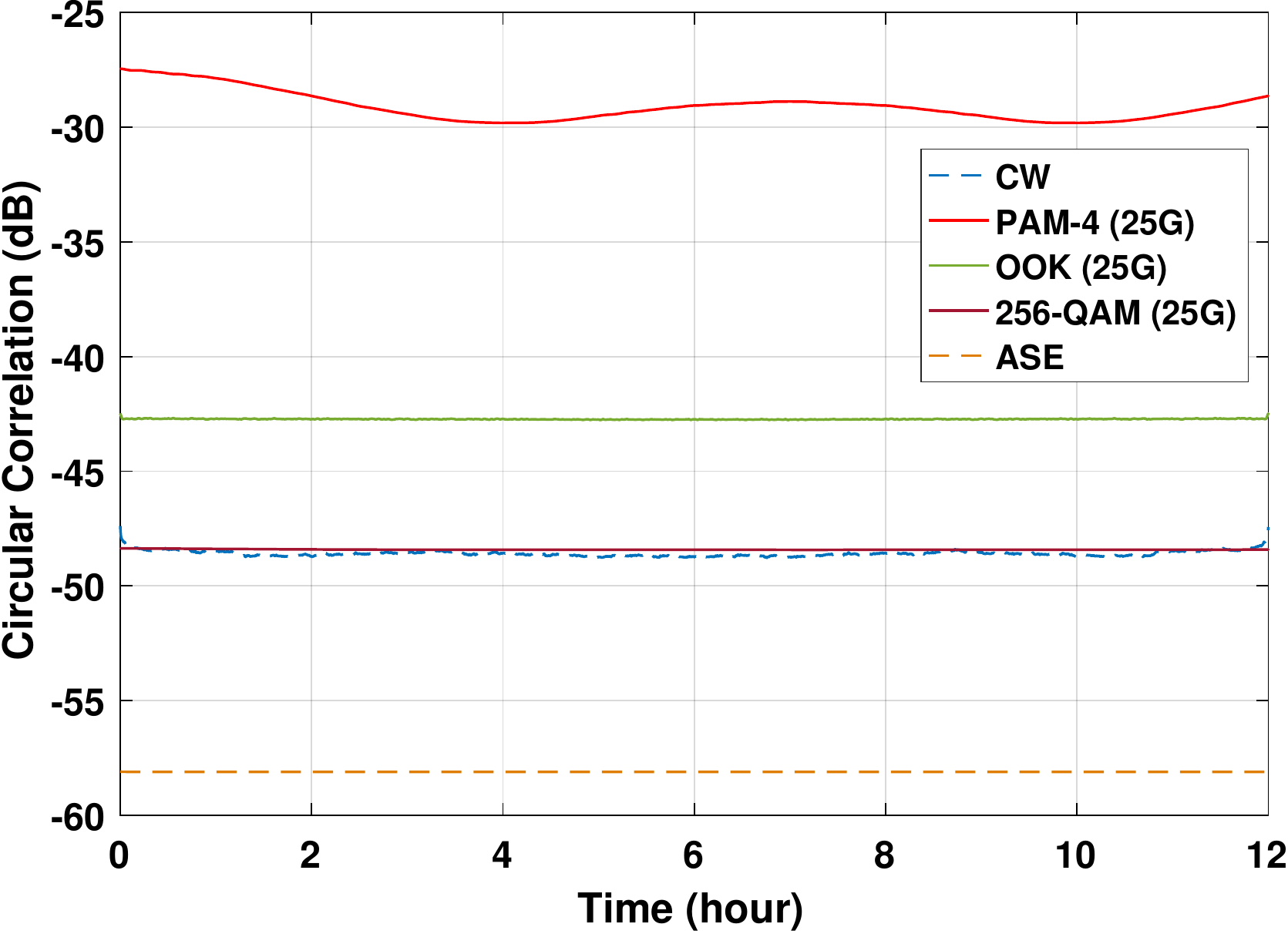}
	\caption{Circular correlation of IC-XT for various signals}
	\label{fig:sc}
\end{minipage}%
\hfill 
\begin{minipage}[t]{\columnwidth}
\centering
\includegraphics[width=0.99\columnwidth]{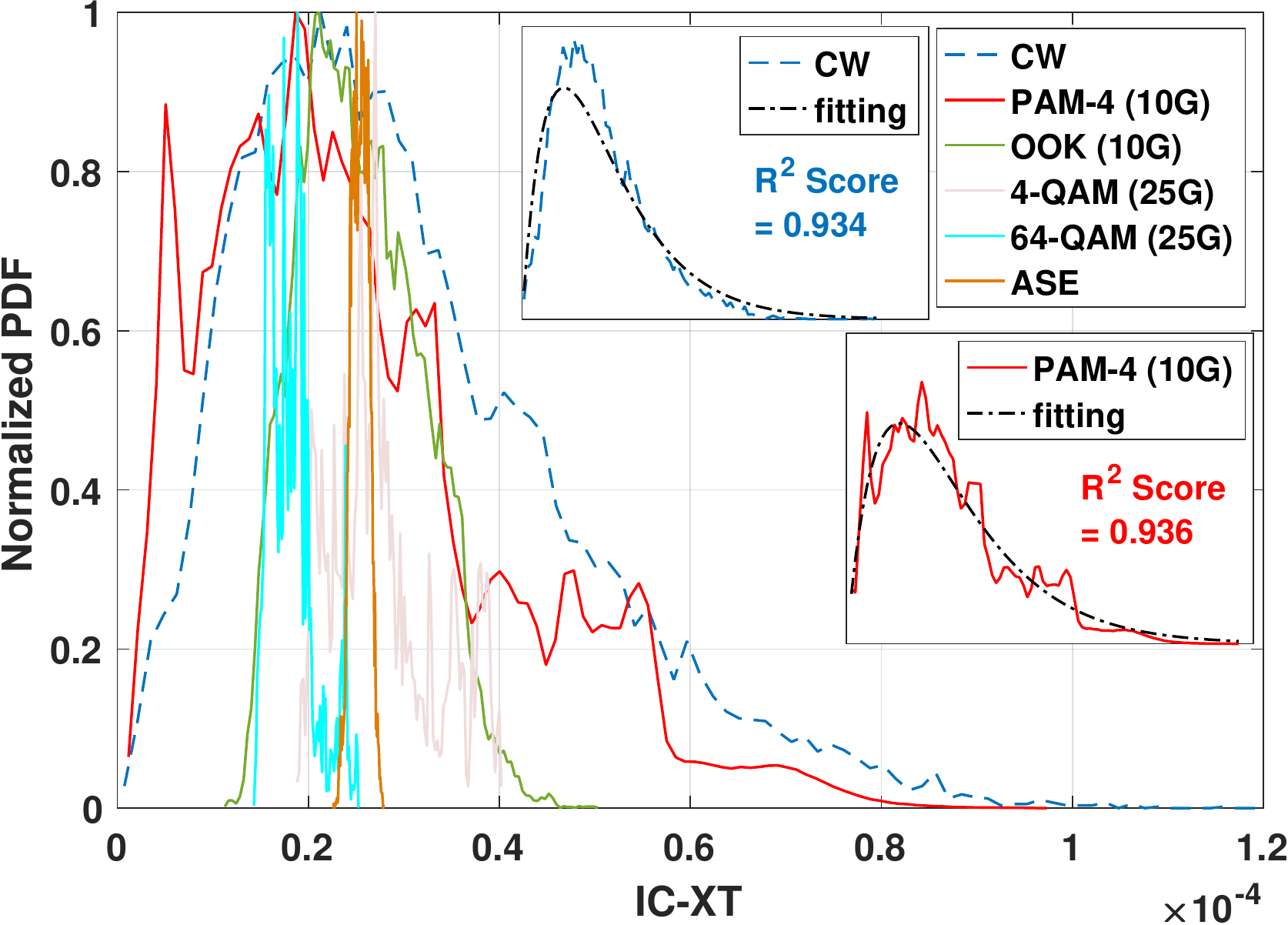}
\caption{Normalized PDF of IC-XT for various signals}
\label{fig:pdf}
\end{minipage}
\end{figure*}

\begin{figure*}[t]
\begin{minipage}[t]{0.66\columnwidth}
\centering
	\includegraphics[width=\columnwidth]{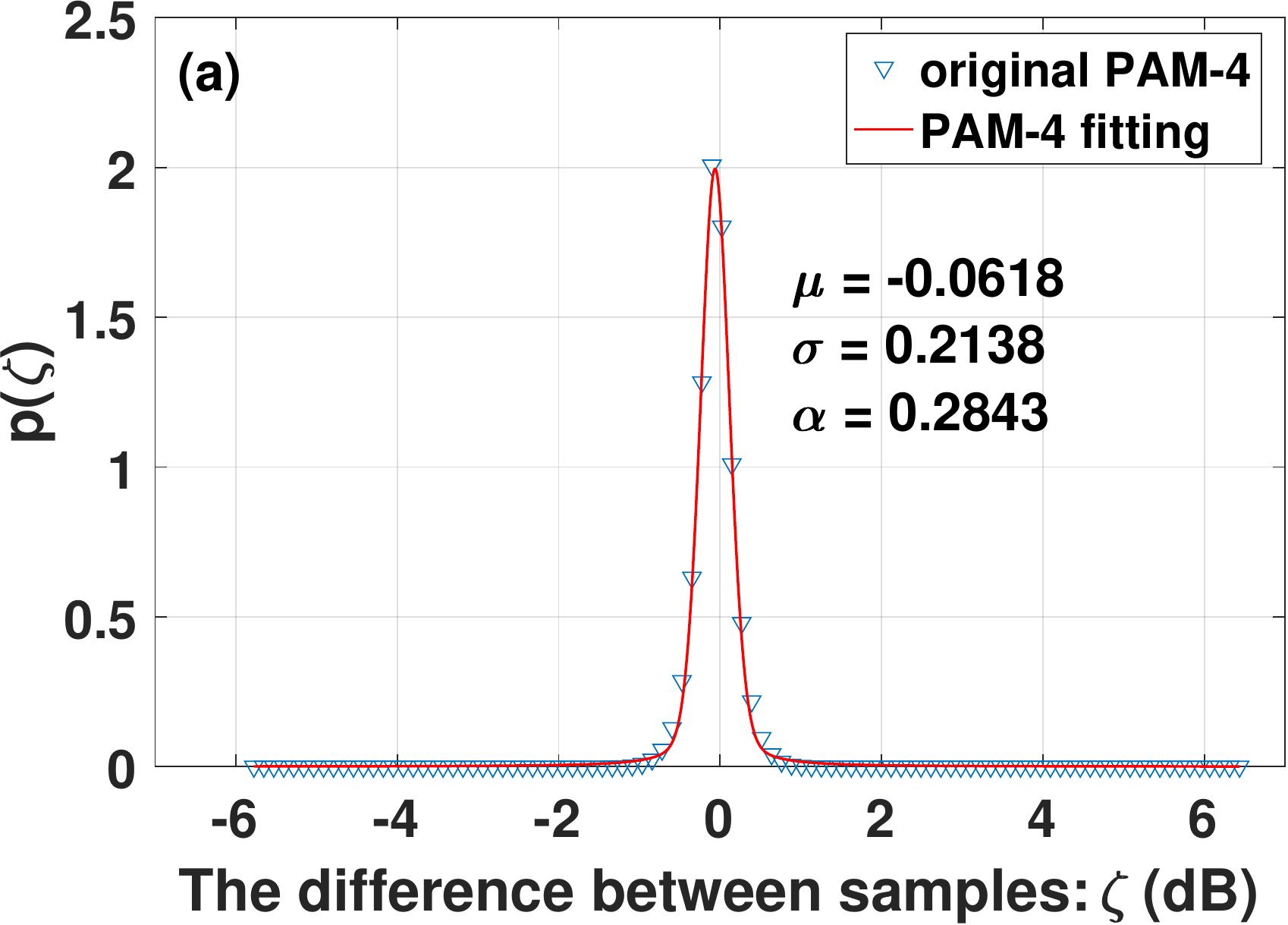}
\end{minipage}%
\hfill 
\begin{minipage}[t]{0.66\columnwidth}
\centering
	\includegraphics[width=\columnwidth]{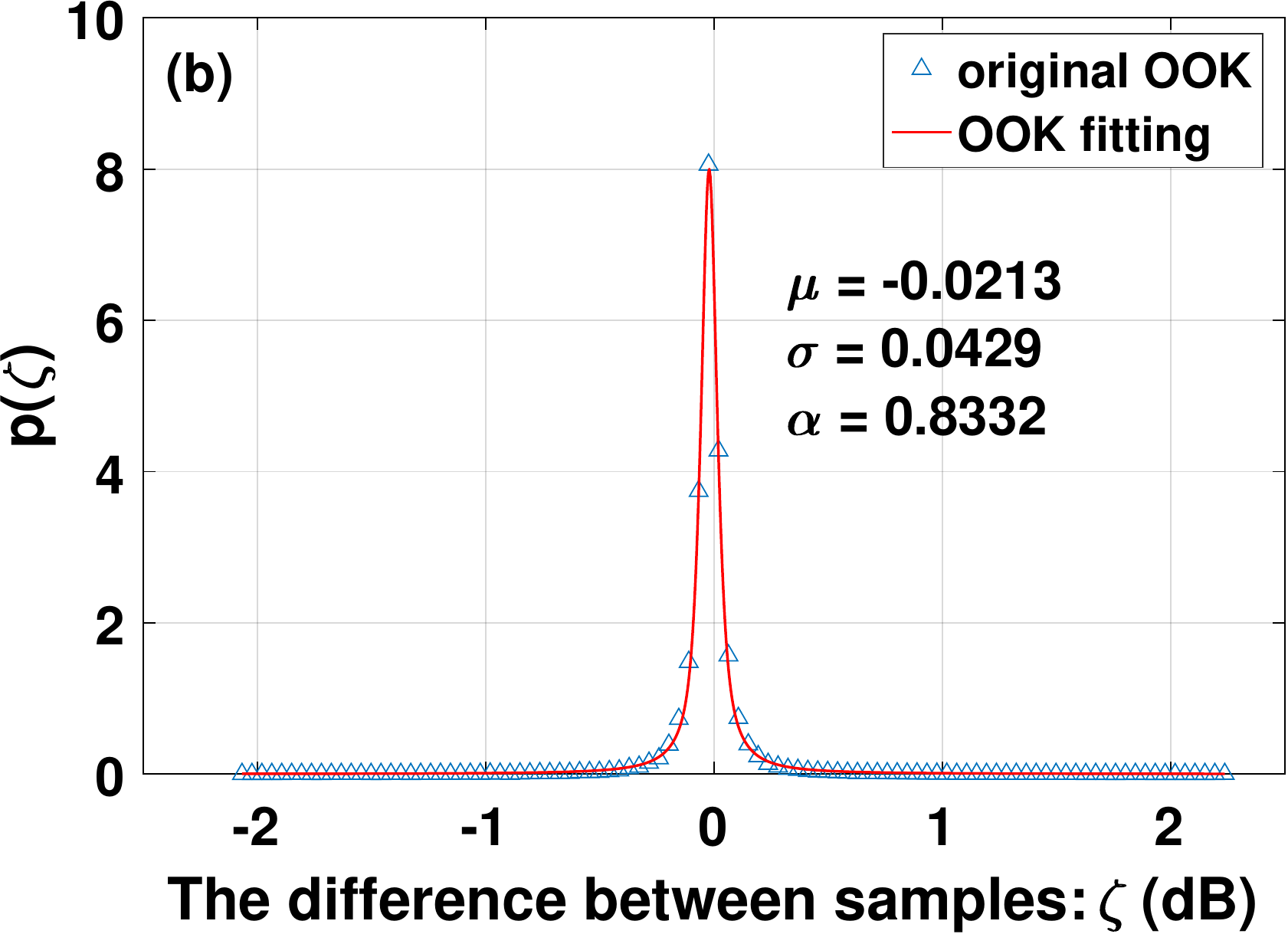}
\end{minipage}
\hfill 
\begin{minipage}[t]{0.66\columnwidth}
\centering
	\includegraphics[width=\columnwidth]{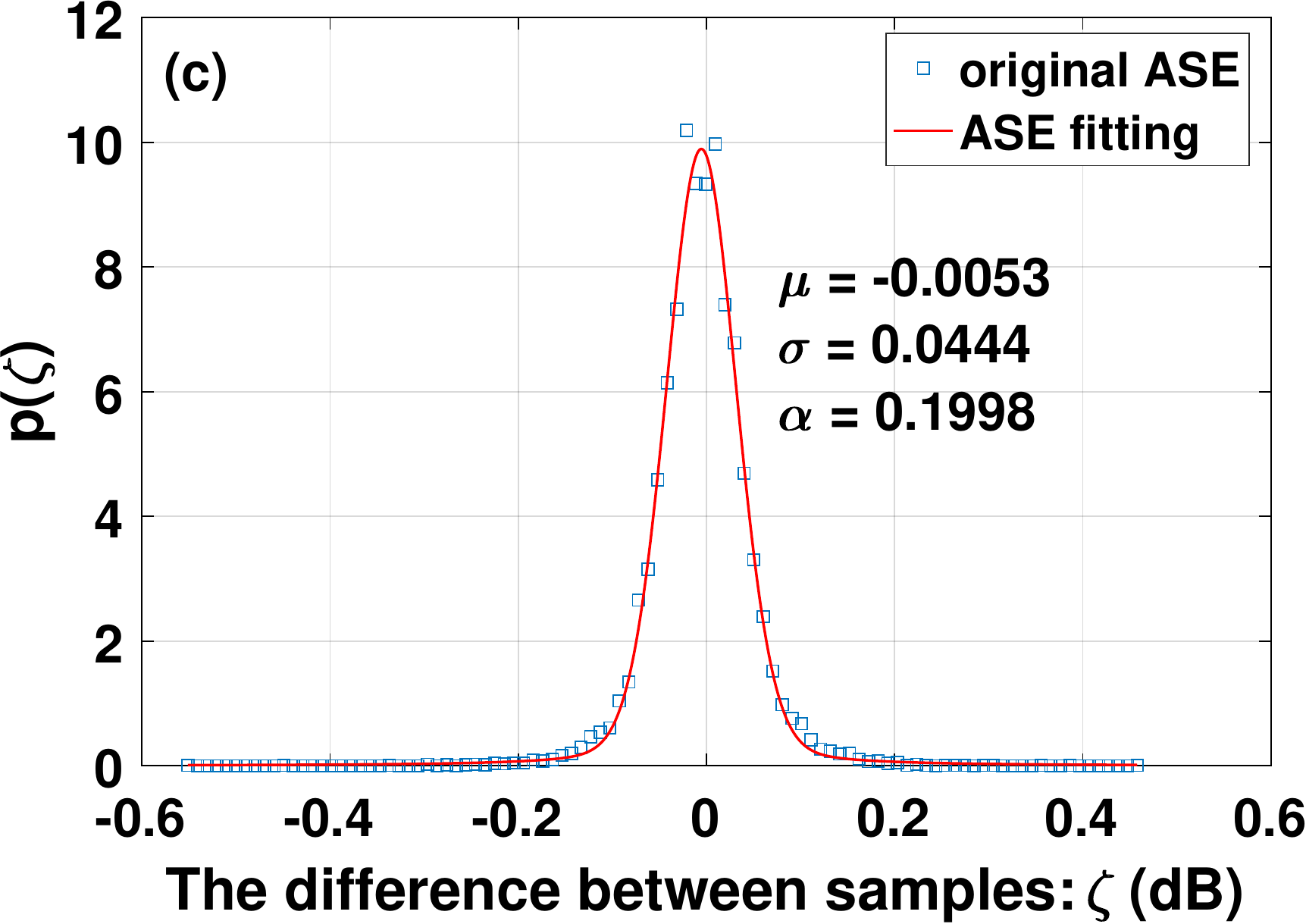}
\end{minipage}
\caption{PDF of IC-XT step for different light signals: a) PAM-4 (25G), b) OOK (25G), and c) ASE}
\label{fig:PVP}
\end{figure*}

\subsection{Analysis of IC-XT step distribution}
\label{sec:profile}

To have a better understanding of the IC-XT, we also investigate the properties of the observed IC-XT data sequence. Firstly, we evaluate the circular correlation, which is a computationally efficient method to measure auto-correlation of a sequence~\cite{Cross_2019}, of the IC-XT signals induced by different signalling sources and the results are presented in Fig.~\ref{fig:sc}. As seen, no matter what kind of signal is launched into the MCF, the circular correlation for IC-XT itself is negligible ($<$-27 dB or 0.002), which indicates that IC-XT sequence is a random process.

According to~\cite{Hayashi_2014}, the distribution of IC-XT follows a Chi-square distribution, and experimental model validation for CW and OOK sources have been done in~\cite{Alves_2019} and~\cite{Luís_2016}, respectively. Figure~\ref{fig:pdf} presents the normalized probability density function (PDF) of the 12-hours IC-XT for different investigated source signals in this research. As seen, IC-XT induced by CW and PAM-4 sources can greatly fit the Chi-square distribution model utilized in \cite{Luís_2016}, presenting 93.4\% and 93.5\% fitting accuracies, respectively, which are evaluated through a $R^2$ score function~\cite{Rowe_2018}. However, for the other sources, the fitness to the distribution reduces accordingly. For example, the $R^2$ score for 10G OOK is 0.401. Especially, for QAM or ASE sources, the observed PDF of IC-XT cannot be fitted to the Chi-square distribution (i.e. the $R^2$ scores become negative). Moreover, it is found that even for the CW source, a Chi-square distribution can only be achieved with a relative long time window (e.g. $>$ 10 hours) and a large number of samples. For instance, when the time window is less than 6 hours, the observed data cannot greatly fit a Chi-square distribution (i.e. the $R^2$ score $<$ 0.8). 

This previously described method (Chi-squared distribution with 4 degrees of freedom) cannot be used to represent OOK, ASE, QAM signals and it requires significant amount of data to accurately fit CW and PAM-4 signals. In addition, the PDF for IC-XT induced by each type of signals is not coherent since when two different portions of the same data sequence were analyzed, the results were very different. For the aforementioned reasons, this method can be considered unreliable for signal statistical analysis, which necessitates the development of a new model. To do so, we propose to investigate the IC-XT as a Pseudo random-walk process based on the reasons: a) IC-XT sequence is a completely random process as aforementioned, b) the analysis in~\cite{Alves_2018} of the auto-correlation extremely resembles the one of a random-walk (cumulative sum of stochastic random variables), for which elements closer to each other in time will be more similar and therefore more correlated, and c) according to the principle of convergence of uncertainties~\cite{Coleman_2009}, both the IC-XT and IC-XT step standard deviations are found to converge to a steady numerical value when the number of samples increases. The model can be expressed as: 
 
\begin{equation}
S_{t+1}  = S_{t} + \zeta \backsim p(\zeta \lvert S_{t})
\end{equation}
Where $S_t$ is the value (i.e. IC-XT) at a certain time and $\zeta$ is a stochastic random variable (i.e. IC-XT step or difference between every adjacent pair of IC-XT samples at a discrete time interval), which follows a particular distribution depending on the value of $S_{t}$. This process reduces the analysis to a single stochastic random variable $\zeta$. For this reason, we model the PDF of $\zeta$, and it is found that it can be expressed as:
 
\begin{equation}
\label{eq:PVP}
p(\zeta)  = \int_{-\infty}^{\infty} p(\zeta \lvert S_{t})p(S_{t})\, d{S_{t} \backsim PVP(\zeta;\mu, \sigma, \alpha)}
\end{equation}

Where PVP stands for the Pseudo-Voigt profile, a numerical approximation of a Voigt profile, which is a convolution between a Cauchy-Lorentz distribution and a Gaussian distribution~\cite{Pagnini_2018,Function}. PVP consists of a weighted sum between a Gaussian and a Cauchy-Lorentz distribution with same mean $\mu$ and differently scaled standard deviation $\sigma$ (scaled version $\sigma_g$), which can be expressed as:
 
\begin{equation}
\begin{split}
PVP(\zeta;\mu, \sigma, \alpha) = \frac{\left(1-\alpha\right)}{\sigma_g \sqrt{2 \pi}}e^{\left[-\left(x-\mu\right)^2/{2{\sigma_g}^2}\right]} \\ + \frac{\alpha}{\pi}\left[\frac{\sigma}{\left(x-\mu\right)^2+\sigma^2}\right]
\end{split}
\end{equation}
In which, {$\sigma_g=\sigma/\sqrt{2 \text{ln} 2}$}. The first part of the equation relates to the Gaussian distribution and the second to the Cauchy-Lorentz distribution, $\alpha$ is the scaling coefficient between the two distributions ($0<\alpha<1$).

\begin{figure*}[t]
\begin{minipage}[t]{\columnwidth}
\centering
	\includegraphics[width=0.99\columnwidth]{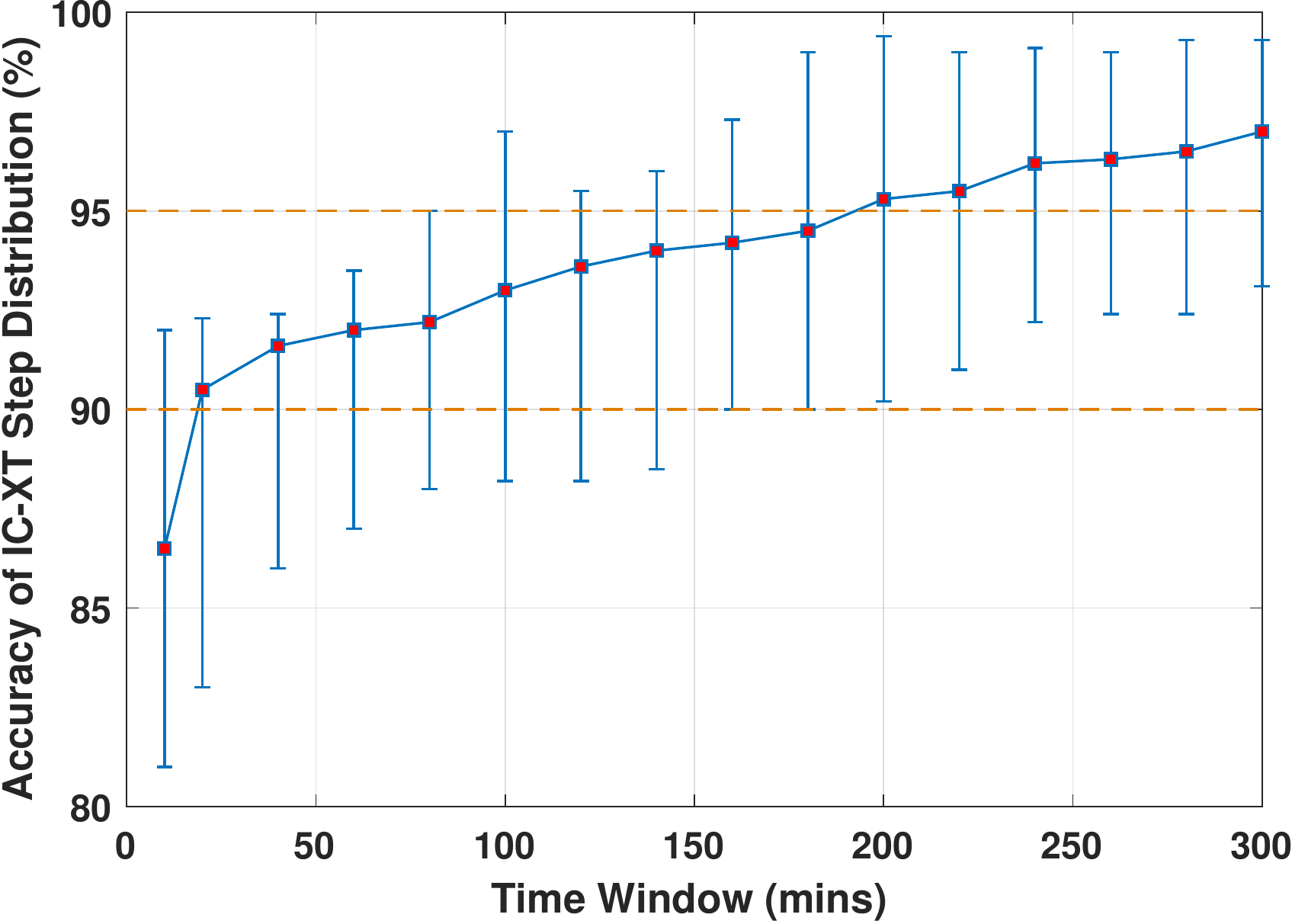}
	\caption{Effect of time window on IC-XT step distribution}
	\label{fig:winpdf}
\end{minipage}%
\hfill 
\begin{minipage}[t]{\columnwidth}
\centering
\includegraphics[width=0.99\columnwidth]{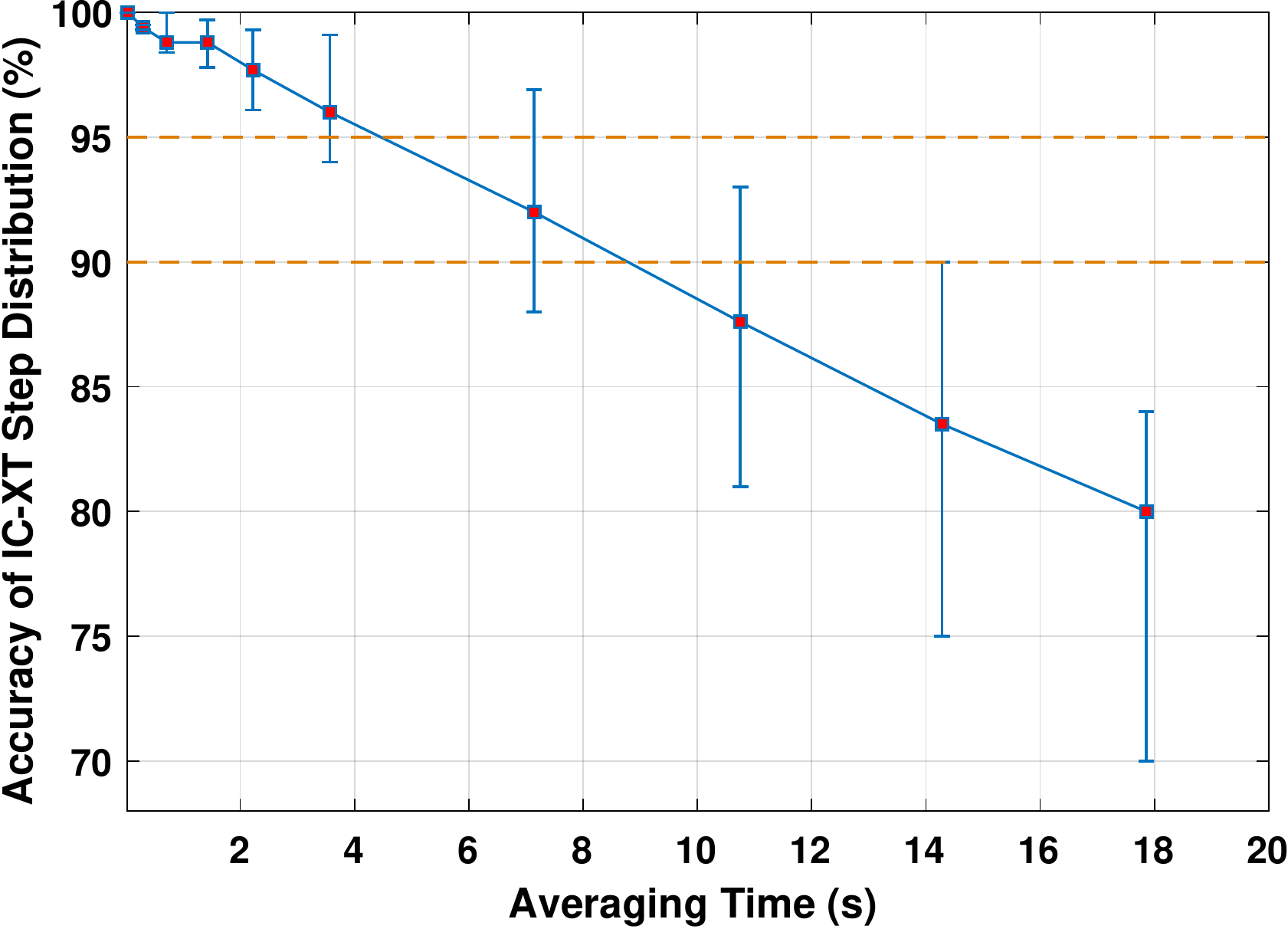}
\caption{Effect of averaging time on IC-XT step distribution}
\label{fig:avgpdf}
\end{minipage}
\end{figure*}

As shown in Fig.~\ref{fig:PVP} and Table~\ref{tab:fit}, the distribution of the observed IC-XT steps for different signalling sources can almost perfectly fit this model ($>$ 99.3\% fitting accuracy). Furthermore, the relationship between the time window and averaging time, and the accuracy of the fitting is also explored and shown in Figs.~\ref{fig:winpdf} and~\ref{fig:avgpdf}. Note that, the distribution accuracy in the figures refers to the similarity of the distribution of current samples to that of the benchmark samples (12-hours time window or 25 ms averaging time), which is also evaluated through the $R^2$ score function. As seen, even with 20~mins observation time, compared with the IC-XT step distribution over 12 hours, around 90\% accuracy is achieved. Moreover, this value can rise to 95\% when the time window is over 80 mins. In addition, Fig.~\ref{fig:avgpdf} shows that to achieve 90\% and 95\% IC-XT step distribution accuracy, 8.5~s and 4~s averaging times are required, respectively, which can be easily realized with the commonly used power meters. Compared with the IC-XT distribution (Chi-square), using the distribution of IC-XT step requires much shorter observation time or less strict averaging time to achieve good fitting accuracy. Taking the IC-XT induced by CW source as an example, when the averaging time is 3.5~s, the minimum time window is 50 minutes to achieve over 90\% accuracy; when the time window equals to 10~mins, the averaging time should be shorter than 0.625~s.  Moreover, this model is suitable for all the investigated sources, including ASE and QAM signals, which can not greatly fit the Chi-square distribution even after 12 hours' measurement, showcasing that low complexity power monitors and short windows can be used to create a crosstalk step distribution that can be used to identify the signal propagated.
 
{\renewcommand{\arraystretch}{1.7}
 \begin{table}[t]
 \centering
 \caption{Fitting coefficients and performance}
\label{tab:fit}
\tabcolsep 0.08in
\begin{tabular}{|c|c|c|c|c|}
\hline
\textbf{Signalling Source} & $\mu$ & $\sigma$ & $\alpha$ & Fitting  Accuracy \\ \hline
\textbf{CW}                & -0.0880    & 0.3712     & 0.8396     & 99.56\%                    \\ \hline
\textbf{OOK}               & -0.0213    & 0.0429     & 0.8332     & 99.85\%                    \\ \hline
\textbf{PAM-4}             & -0.0618    & 0.2138     & 0.2843     & 99.74\%                    \\ \hline
\textbf{256-QAM}           & 0.3675     & 0.0044     & -0.0016    & 99.87\%                    \\ \hline
\textbf{ASE}               & -0.0053    & 0.0444     & 0.1998     & 99.33\%                    \\ \hline
\end{tabular}
\end{table}}

\section{Conclusion}
\label{sec:con}

{\renewcommand{\arraystretch}{1.7}
\begin{table*}[t]
\centering
\caption{IC-XT dependence on the investigated parameters and static/dynamic IC-XT for various signalling sources (1550nm, 25GBaud)}
\label{tab:coeff}
\tabcolsep 0.016in
\begin{tabular}{|c|c|c|c|c|c|c|c|c|}
\hline
\textbf{Coefficient} & \multicolumn{2}{c|}{\textbf{Temperature}} & \textbf{Wavelength} & \textbf{Baud Rate ($x$)} & \multicolumn{3}{c|}{\textbf{PRBS Length ($2^i-1$)}} & \textbf{Excited Cores} \\ \hline
\textbf{Wavelength} & \multicolumn{2}{c|}{1550 nm}& 1480-1630 nm & 1550 nm  & \multicolumn{3}{c|}{1550 nm} & 1550 nm \\ \hline
\textbf{\begin{tabular}[c]{@{}c@{}}Signalling\\ Source\end{tabular}} & ASE & 25G-OOK & 25G-OOK & \begin{tabular}[c]{@{}c@{}}OOK/PAM-4/4-,\\  16-, 64-, 256-QAM\end{tabular} & \multicolumn{3}{c|}{25G-OOK}& 25G-OOK\\ \hline
\textbf{Temperature} & 20-50 $^\circ$C & 20-50 $^\circ$C & 20-50 $^\circ$C & 23 $^\circ$C & 30 $^\circ$C & 40 $^\circ$C & 50 $^\circ$C & 23 $^\circ$C \\ \hline
\textbf{Static IC-XT}   & 0.05 dB/K  & 0.13 dB/K    & 0.113 dB/nm  & independent  & $-1.7log_2i-41.4$ & $-1.9log_2i-40$ & $-2.6log_2i-36.7$ & \multirow{3}{*}{\begin{tabular}[c]{@{}c@{}} fluctuation speed \\  increases by a factor \\ of 7.4 per core \end{tabular}} \\ \cline{1-8}
\textbf{\begin{tabular}[c]{@{}c@{}}Dynamic\\ IC-XT \end{tabular}}  & -0.01 dB/K & -0.09  dB/K & independent & \begin{tabular}[c]{@{}c@{}}QAM: \\${1.429*0.977^x}$\end{tabular} & -  & - & - & \\ \hline\hline

\textbf{Signalling Source}  & \textbf{CW} & \textbf{ASE} & \textbf{OOK} & \textbf{PAM-4} & \textbf{4-QAM} & \textbf{16-QAM} & \textbf{64-QAM} & \textbf{256-QAM} \\ \hline
\textbf{Static IC-XT}  & -46.07 dB & -45.95 dB       & -45.92 dB       & -46.9 dB          & -45.63 dB         & -46.84 dB          & -47.43 dB          & -48.61 dB  \\ \hline
\textbf{Dynamic IC-XT} & 22.58 dB & 0.92 dB       & 6.50 dB          & 12.38 dB          & 3.30 dB            & 2.22 dB            & 2.50 dB             & 1.80 dB  \\ 
\hline
\end{tabular}
\end{table*}}

We conducted a comprehensive investigation on the behaviour of the static and dynamic IC-XT with an 8-core TA-MCF considering the impact of the properties of the stimulating optical signals, such as modulation format, baud rate, temperature, PRBS length, operating wavelength, and the number of excited cores. The coefficients of IC-XT dependence on these parameters are characterized and summarized in Table~\ref{tab:coeff}. In terms of static IC-XT, it is proportional to both temperature and wavelength while inversely proportional to the PRBS length and the order when \emph{m}-ary QAM signals are utilized, signifying the importance of modulation formats, temperature and PRBS length on IC-XT measurements. In contrast, the dynamic IC-XT is inversely proportional to the temperature, baud rate whilst independent on the wavelength. Comparing to intensity modulated signals, i.e. OOK and PAM-4, higher IC-XT stability was obtained when QAM modulation was adopted. The speed of IC-XT fluctuation increases as the number of excited cores increases. Furthermore, the experimental results on wavelength-dependent static IC-XT and power distribution of the time-dependent IC-XT accurately fit the theoretical estimations, validating the effectiveness of the existing analytical models. At last, the exploration of the effects from averaging time and time window on IC-XT accuracy can significantly serve as a reference for the experimental measurement of IC-XT in the labs and in practical applications, while the investigation on the distribution of IC-XT step can serve future machine learning based IC-XT classification. It is expected that the coefficients presented in Table~\ref{tab:coeff} as well as the sophisticated understanding and accurate measurements of IC-XT levels demonstrated in this paper can benefit the design of MCF-based data centers, metro networks or telecommunication systems greatly.


%



\section*{Acknowledgment}

This work is partly supported by EPSRC TRANSNET grant EP/R035342/1, the Royal Academy of Engineering fellowship to Dr L Galdino and EPSRC studentship to E Sillekens. We would also like to thank Santec and Oclaro for the provision of the necessary equipment.

\section*{Data and measurement access}

A comprehensive subset of the measurement generated by the experiments performed in this manuscript is publicly available at \cite{Dataset:measurements}.

\ifCLASSOPTIONcaptionsoff
  \newpage
\fi



\bibliographystyle{IEEEtran}
\bibliography{references}

\end{document}